\documentclass[twocolumn,showpacs,preprintnumbers,amsmath,amssymb,showkeys]{revtex4}

\usepackage{lineno}
\usepackage{graphicx} 
\usepackage{dcolumn} 
\usepackage{bm} 
\usepackage[perpage]{footmisc}
\usepackage{amsmath,amsfonts,amssymb,wasysym,ifsym}
\usepackage{color}
\usepackage{MnSymbol}
\usepackage{enumitem}

\newcommand \leff{$\mathcal{L}_{\mathrm{eff}}$}
\newcommand \ly{$\mathcal{L}_{\mathrm{y}}$}
\newcommand \qy{$\mathcal{Q}_{\mathrm{y}}$}
\newcommand \kevnr{$\mathrm{keV_{nr}}$}
\newcommand \kevee{$\mathrm{keV_{ee}}$}

\begin{document}

\title{A New Analysis Method for WIMP searches with Dual-Phase Liquid Xe TPCs} 


\author{K.~Arisaka}
\author{P.~Beltrame\footnote{Corresponding author \\
{\it Email address}: pbeltrame@physics.ucla.edu (P. Beltrame)}}
\author{C.~Ghag}
\author{K.~Lung}
\author{P.R.~Scovell}

\address{Department of Physics \& Astronomy, University of California, Los Angeles, CA, 90095, USA}
\date{\today} 

\begin{abstract}

\noindent A new data analysis method based on physical observables for WIMP dark matter searches with noble liquid Xe dual-phase TPCs is presented. Traditionally, the nuclear recoil energy from a scatter in the liquid target has been estimated by means of the initial prompt scintillation light (S1) produced at the interaction vertex.  The ionization charge (C2), or its secondary scintillation (S2), is combined with the primary scintillation in log$_{10}$(S2/S1) vs. S1 only as a discrimination parameter against electron recoil background. Arguments in favor of C2 as the more reliable nuclear recoil energy estimator than S1 are presented. The new phase space of log$_{10}$(S1/C2) vs. C2 is introduced as more efficient for nuclear recoil acceptance and exhibiting superior energy resolution.  This is achieved without compromising the discrimination power of the LXe TPC, nor its 3D event reconstruction and fiducialization capability, as is the case for analyses that exploit only the ionization channel. Finally, the concept of two independent energy estimators for background rejection is presented: E2 as the primary (based on C2) and E1 as the secondary (based on S1).  log$_{10}$(E1/E2) vs. E2 is shown to be the most appropriate phase space in which to evaluate WIMP signal candidates.

\end{abstract}


\maketitle

\section{INTRODUCTION}

Considerable astronomical and cosmological evidence supports a self-consistent $\Lambda$CDM model of the Universe where approximately 23\% is in the form of dark matter. Discovery of the nature of this dark matter is recognized as one of the greatest contemporary challenges in science, fundamental to our understanding of the Universe. The most compelling candidates for dark matter are Weakly Interacting Massive Particles (WIMPs) that arise naturally in several models of physics beyond the Standard Model~\cite{Steigman:1985,Jungmann:1996}.

WIMPs may be directly detected through the energy deposited by a recoiling nucleus following a rare WIMP scatter with standard matter.  Given the kinematics, the recoil would generate an extremely small signal of $<$100 \kevnr\ (keV nuclear recoil) requiring detectors to exhibit good sensitivity (high detection efficiency) at low energies (down to a few \kevnr) with good energy resolution to determine the WIMP energy spectrum and estimate the WIMP mass. Simultaneously such an ideal detector must have good rejection power for background events from $\gamma$-rays. Despite this, the experimental sensitivities of direct searches are now entering into the theoretically favored parameter space and a positive detection may be imminent from current or next generation devices. Two-phase noble liquid time projection chambers (TPCs) offer particularly attractive prospects for dark matter detection through excellent background rejection capability, 3D position sensitivity allowing definition of a fiducial volume, and cost-effective scalability~\cite{Angle:2008,Alner:2007,Amsler:2008,Wright:2011,Akimov:2012,brunetti:2005,Aprile:2011instr,Aprile:2011Run08,Akimov:2011,Lebedenko:2009}. Liquid xenon (LXe) in particular offers perhaps the most promising prospect for unambiguous detection: it is intrinsically radio-pure, has the capability of powerful self-shielding~\cite{Boulay:2008,Sekiya:2010,Gomez:2009}, and is sensitive to low energy nuclear recoils. The background rejection power of two-phase TPCs comes from the ability to record both direct scintillation light (S1) and electroluminescence from ionization (S2) of the target following an energy deposition.  The ratio of the signal strength in these channels differs for electron and nuclear recoil interactions, allowing efficient discrimination between incident particle species. 

Traditionally the nuclear recoil energy scale for LXe TPCs with photomultiplier (PMT) read-out is set with the prompt primary scintillation channel S1. However, this channel is dominated by statistical fluctuations in signal generation and detection that ultimately preclude a direct mapping from observed S1 to recoil energy. Although sufficient for setting limits on WIMP-nucleon cross-sections where no signal is observed, the S1 energy scale severely limits the ability of a LXe TPC to test any possible observed signal excess against the WIMP hypothesis through analysis of the energy spectrum and event distribution. Furthermore, the requirement of S1 signal across multiple PMTs imposes a relatively high energy threshold.  In attempts to circumvent such limitations, an alternative approach is based on the ionization S2 channel alone. Although allowing a considerably lower threshold with sensitivity down to single electrons and a superior energy resolution due to electroluminescence gain in the gas phase, the ability to discriminate background is lost and, in the absence of a prompt S1, the 3D vertex identification is compromised. Furthermore, significant background (from, for {\em e.g.}, VUV photoionization \cite{Santos:2011,Edwards:2008}) may limit the sensitivity of such S2-only analyses.

In an effort to retain the discrimination power and fiducialization capability of the LXe TPC whilst simultaneously exploiting the low energy threshold and superior resolution afforded to the ionization channel, the adoption of a new data analysis method is proposed here. It is shown that the traditional representation of data in log$_{10}$(S2/S1) vs. S1 phase space can be replaced with log$_{10}$(S1/C2) vs. C2 (where C2 is the ionization yield) with no increase in background but with significant improvement in energy resolution allowing reduction in the uncertainty in conversion to true recoil energy. This new technique, whilst retaining all the benefits of the LXe TPC technology such as self-shielding through fiducial definition and particle species discrimination, allows the direct conversion to an energy on an event-by-event basis. Proposed combined (scintillation and ionization) nuclear recoil energy scales \cite{Sorensen:2011,Lindner:2011,Shutt:2007} are still affected by the poor resolution in the S1 channel for low energy recoils, whilst C2 represents a more appropriate energy scale.

This article is organized as follows.  In Section~\ref{sec:c2intro} the traditional energy estimator, S1, and the ionization channel, C2, are contrasted and modeled.  Limitations of S1 as the energy scale are highlighted, and C2 shown to provide superior energy resolution and lower detector energy threshold for Xe. In Section~\ref{sec:phase_spaces} the log$_{10}$(S2/S1) vs. S1 and log$_{10}$(S1/C2) vs. C2 phase spaces are examined with the aid of a simple yet realistic model and monoenergetic neutrons. This illustrates the inefficiency in signal acceptance, especially at higher energies, and misrepresentation of energy in the traditional phase space. Appropriate signal acceptance regions, with clear energy thresholds that need not exploit smearing of sub-threshold energy depositions into signal regions, are defined in the log$_{10}$(S1/C2) vs. C2 phase space. 
The log$_{10}$(E1/E2) vs. E2 phase-space is then introduced - these quantities are no longer detector specific and accessible as recoil energy for direct comparisons with other technologies. Finally, the response of the model detector to WIMPs is examined in light of the new phase-space.

\section{Nuclear Recoil Energy Scale}
\label{sec:c2intro}

\subsection{S1 and C2 as nuclear recoil energy estimators}

\begin{figure}[htb!]
\begin{center}
\includegraphics[width=.5\textwidth]{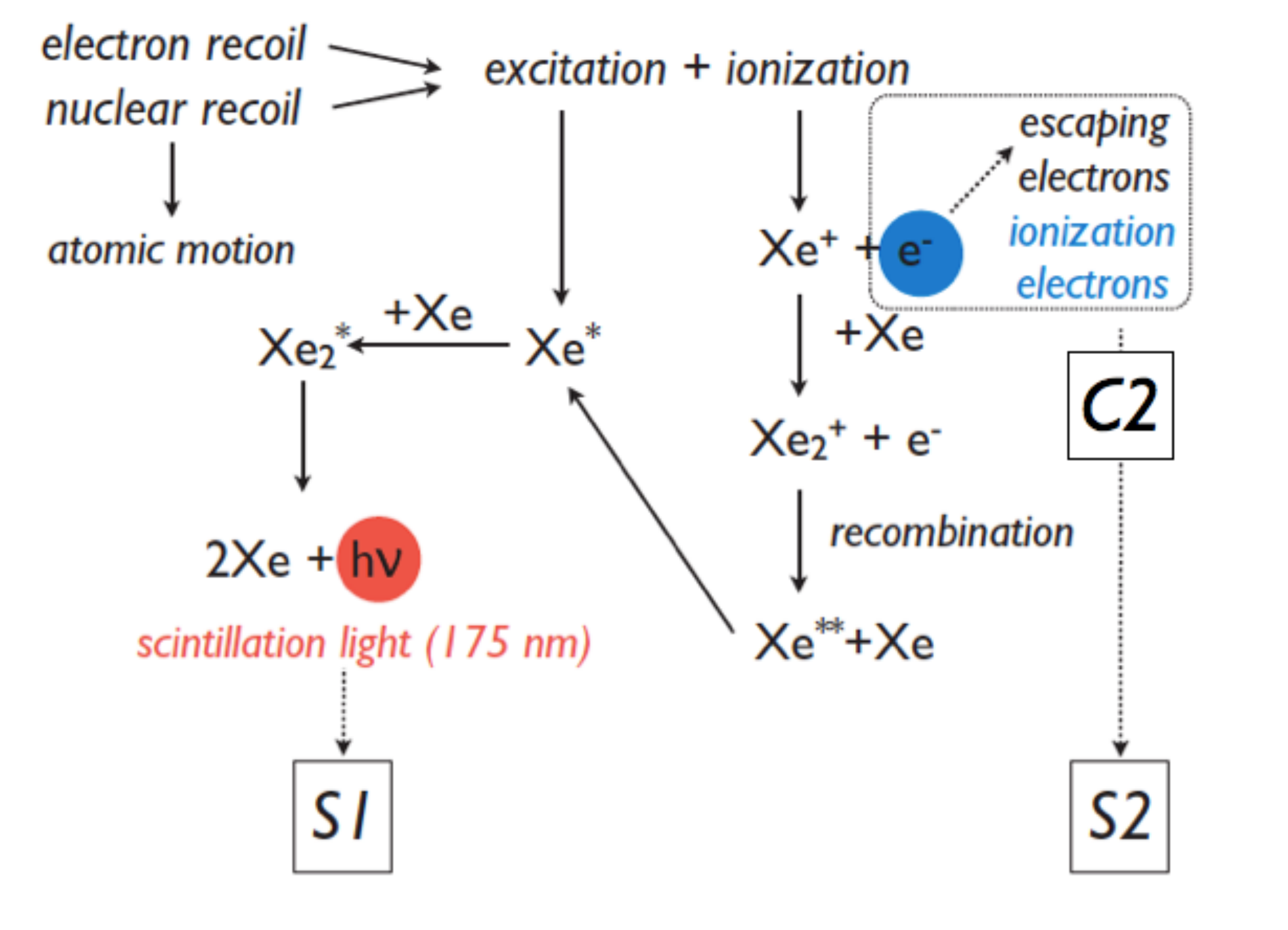}
\caption{\small{Diagram of the production of excitation (S1) and ionization (C2, and subsequent S2) signals in 2-phase liquid/gas detectors (adapted from~\cite{manzur:2010}).}}
\label{fig:s1s2generation}
\end{center}
\end{figure}

Fig.~\ref{fig:s1s2generation} illustrates the mechanism by which two-phase LXe TPCs record both the so-called S1 and S2 signals following an interaction in the target volume. Energy deposition in LXe leaves atoms in both excited and ionized states. Excited Xe atoms combine with un-excited Xe atoms in a matter of a few picoseconds to form an excited Xe$_{2}^{*}$ molecule. The de-excitation of this molecule releases ultraviolet photons. Simultaneously, ionized Xe atoms combine with non-ionized atoms to form Xe$_{2}^{+}$ molecules. In the absence of an electric field, these recombine with electrons to form an excited  Xe$^{**}$ atom that emits ultraviolet photons through de-excitation (S1). In the presence of an electric field, not all the free electrons are able to recombine with the ionized Xe molecule and are drifted up to the liquid/gas interface. Under a strong enough electric field, these free electrons (C2) are extracted from the liquid surface and travel through the gas phase, causing electroluminescence and giving a secondary scintillation signal (S2). The conversion from C2 to S2 is detector dependent - related to the gas gap and electric field configuration of the detector - it is simply the number of photoelectrons per electron extracted into the gas phase. For this reason C2, as the more meaningful parameter, is used as the primary observable rather than S2 throughout this article.

\begin{figure}[htb!]
\begin{center}
\includegraphics[width=.5\textwidth]{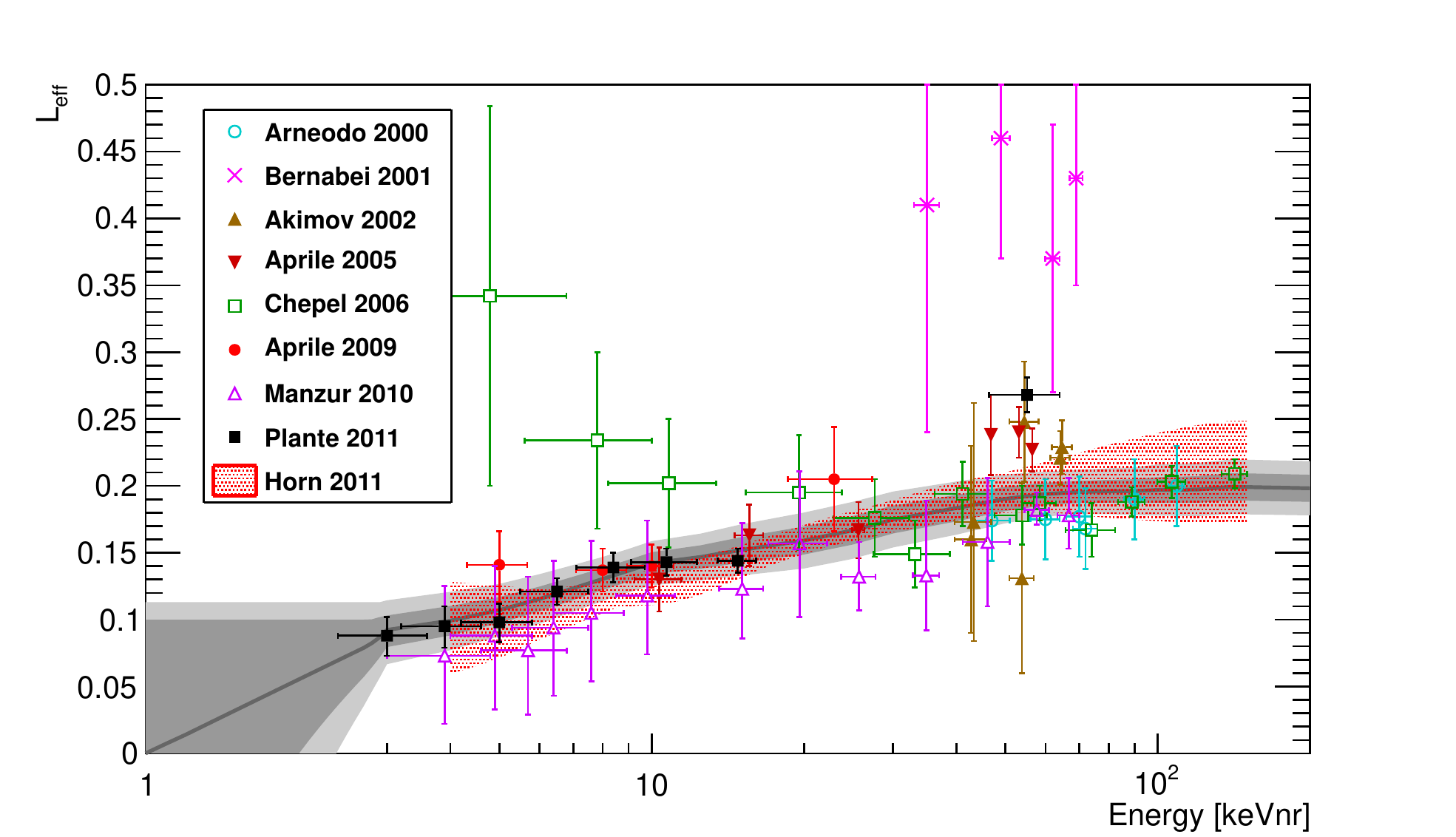}
\caption{\small{Collection of all direct measurements of \leff\ to date~\cite{arneodo:2000,bernabei:2001,akimov:2002,aprile:2009,Plante:2011}. In this diagram, the solid grey regions represent a Maximum Likelihood fit to all direct measurements and the associated 1$\sigma$ and 2$\sigma$ confidence bands; this parameterization is adopted in this work and is taken from \cite{Aprile:2011Run08}. \leff\ is extrapolated to zero at 1~\kevnr. In addition, a contemporaneous indirect (data vs. Monte Carlo) measurement \cite{Horn:2011} is shown to be in good agreement with direct measurements.}}
\label{fig:leff}
\end{center}
\end{figure}

Using S1 as the primary nuclear recoil energy estimator appears to suffer from three fundamental difficulties:
\begin{enumerate}
\item The S1 signal is extremely small: typically a fraction of single photoelectron (PE) per keV at energies below 10~\kevnr. 
\item Conversion from S1 to nuclear recoil energy, $E_{\mathrm{nr}}$ depends on several parameters, given in Eq.~\ref{leffeq}:

\begin{equation}\label{leffeq}
E_{\mathrm{nr}} = \frac{\mathrm{S1}}{\mathcal{L}_\mathrm{y}}  \frac{1}{\mathcal{L}_{\mathrm{eff}}}    \frac{S_{\mathrm{nr}}}{S_{\mathrm{ee}}},
\end{equation}

where $S_{\mathrm{nr}}$ and $S_{\mathrm{ee}}$ describe the electric field suppression factors for nuclear and electron recoils, respectively, 
\ly\ is the detector light yield, and \leff\ is the scintillation yield for nuclear recoils.

First the absolute \ly\ calibration point is given by 122 \kevee\ (keV electron equivalent) $^{57}$Co $\gamma$-rays. Second, the difference between field dependent detector response to $\gamma$-rays and nuclear recoils must be accounted for ($S_{\mathrm{nr}}$ and $S_{\mathrm{ee}}$). Finally, the energy dependence of the scintillation yield for nuclear recoils is corrected by \leff\ (shown in Fig.~\ref{fig:leff}).  

\item \leff\ is suppressed at lower energies, resulting in poorer energy measurements due to a lack of photoelectron statistics.
\end{enumerate}

\begin{figure}[htb!]
\begin{center}
\includegraphics[width=.5\textwidth]{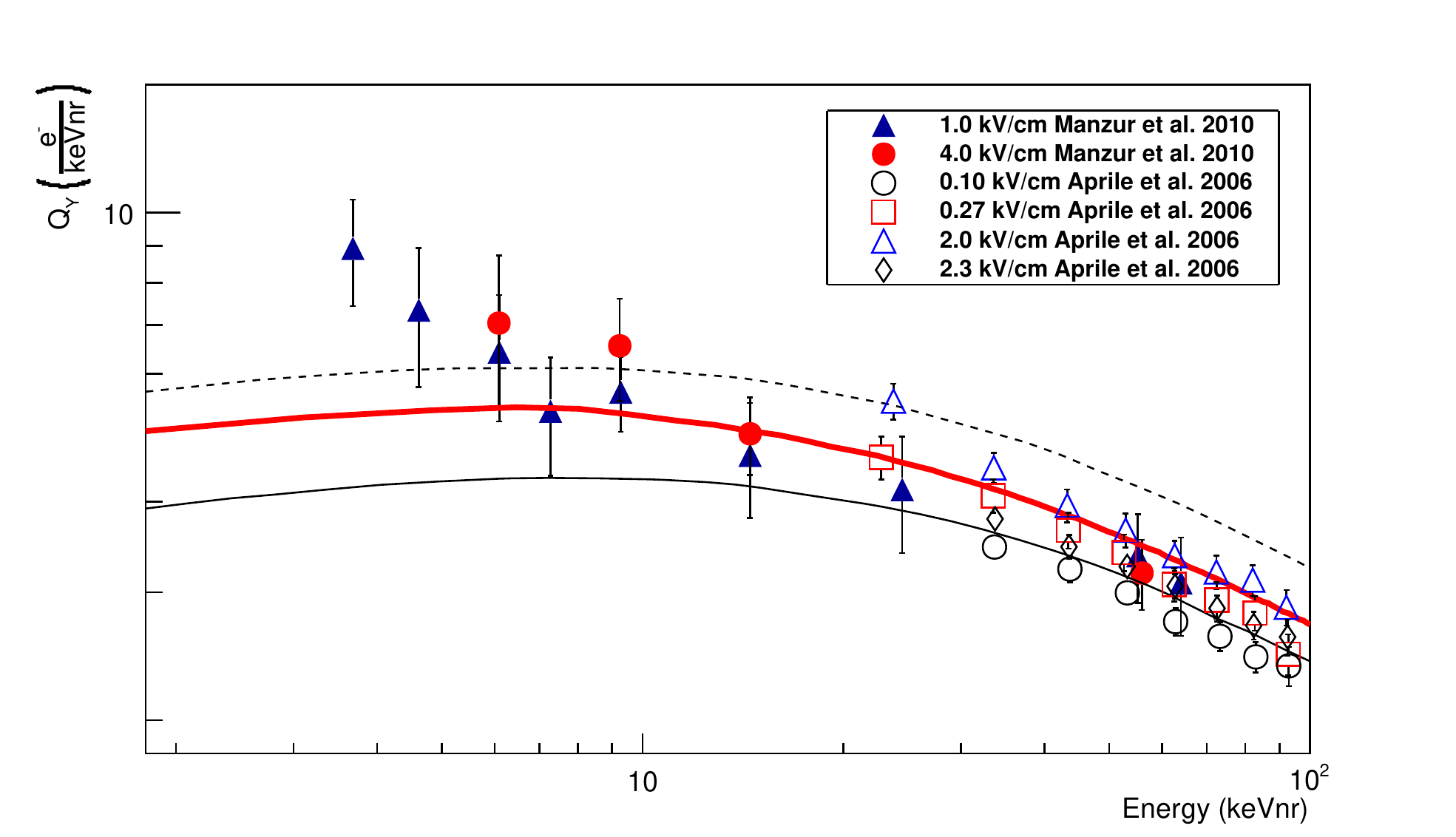}
\caption{\small{Charge yield (\qy) as measured directly at several electric fields \cite{manzur:2010,aprile:2006}. Also included are theoretical \qy\ curves derived using stopping powers as calculated by Hitachi \cite{hitachi:2005} (dashed curve) and Lindhard \cite{lindhard:1963} (solid curve) and extracted from \cite{Angle:2011}. For the purposes of this paper, a conservative \qy\ that falls between the Hitachi and Lindhard measurements is adopted (red curve).}}
\label{fig:qy}
\end{center}
\end{figure}

It is well known that such extrapolation (from 122 \kevee\ $\gamma$-rays to a few \kevnr\ neutrons) and conversions tend to propagate large systematic errors at each stage. In other words, this is a so-called `piece-by-piece' calibration, in opposition to ideal `end-to-end' calibrations that tend to cancel out systematic errors.
The investigation presented here indicates that the ionization yield, C2, can act as a much improved energy estimator for the following three reasons:
\begin{enumerate}
\item The yield of C2 is an order of magnitude larger: typically $\sim$5 electrons per \kevnr. 
\item Conversion from C2 to nuclear recoil energy is straightforward:  
\begin{equation}\label{c2qy}
E_{\mathrm{nr}} = \frac{\mathrm{C2}}{\mathcal{Q}_\mathrm{y}}, 
\end{equation}
depending only on the one parameter \qy, the charge yield, such that the systematic error in the conversion is reduced. The energy dependence of \qy\ is shown in Fig.~\ref{fig:qy}.
\item Theoretical models and dedicated measurements indicate that the value of \qy\ increases with decreasing energy, meaning that even at low energy, large C2 signals are expected.
\end{enumerate}

\subsection{Modeling S1 and C2 signals}

In order to perform a further systematic comparison, a model is constructed using Monte Carlo simulations. The underlying, well-motivated assumptions \cite{Horn:2011,Plante:2011,hitachi:2005,lindhard:1963,Sorensen:2010, Angle:2011} are as follows:
\begin{itemize} 
\item The energy dependence of \leff\ and \qy\ is as shown in Figs.~\ref{fig:leff} \&~\ref{fig:qy}, respectively.    
\item \ly = 3 PE/\kevee\ at 122 \kevee.
\item $S_{\mathrm{nr}}$ and $S_{\mathrm{ee}}$ are chosen assuming an electric field of 0.5 kV/cm.
\item For conversion from S2 to C2, the secondary scintillation gain in a gas phase is assumed to be 20 ({\em i.e.}, S2 = C2~$\times$~20), with negligible variation across the surface of any fiducial volume.    
\item Poisson distribution for S1 in terms of the number of photoelectrons.
\item Poisson distribution for C2 in terms of the number of free electrons from initial ionization.  
\end{itemize}

At this stage new terminology may be introduced: `E1' as the energy estimated by S1, and `E2' as the energy estimated by C2.  The dominant source of the energy mis-measurement is the Poisson fluctuations in S1 in the case of E1, and in C2 for the case of E2.  It should be noted that any other factors such as the secondary scintillation process from C2 to S2 have a negligible contribution.  E2 by itself represents a better estimate of the energy than the combination of E1 and E2 (as adopted for electromagnetic interactions of $\gg$10 \kevee) due to the poor photoelectron statistics of S1.

\begin{figure}[htb!]
\begin{center}
\includegraphics[width=.5\textwidth]{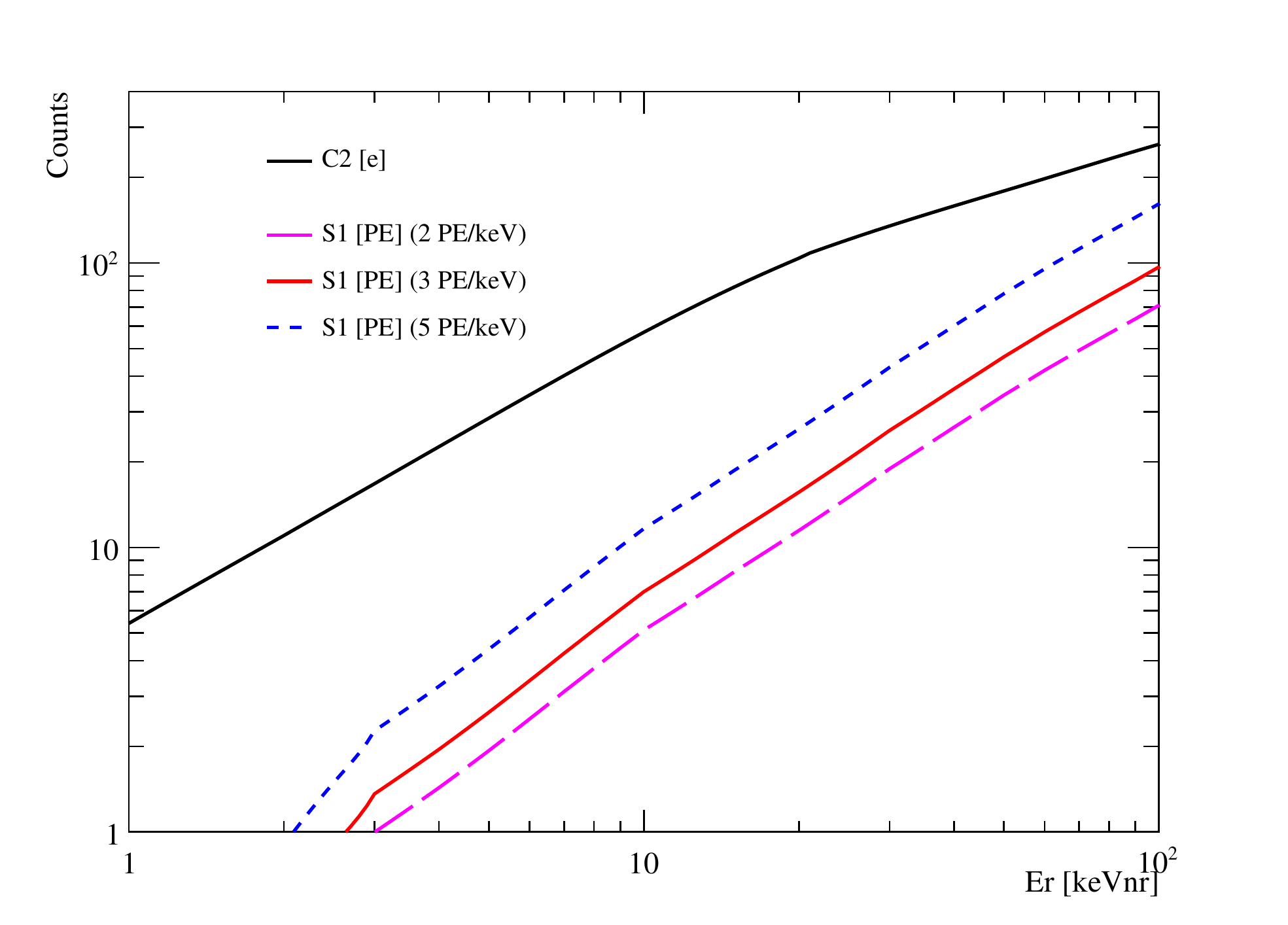}
\caption{\small{S1, for three different detector light yields, and C2 observables as a function of nuclear recoil energy. S1 is shown in units of PE, whereas C2 is in electrons ({\em i.e.}, prior to amplification to S2).}}
\label{fig:s1c2energy}
\end{center}
\end{figure}

\begin{figure}[htb!]
\begin{center}
\includegraphics[width=.5\textwidth]{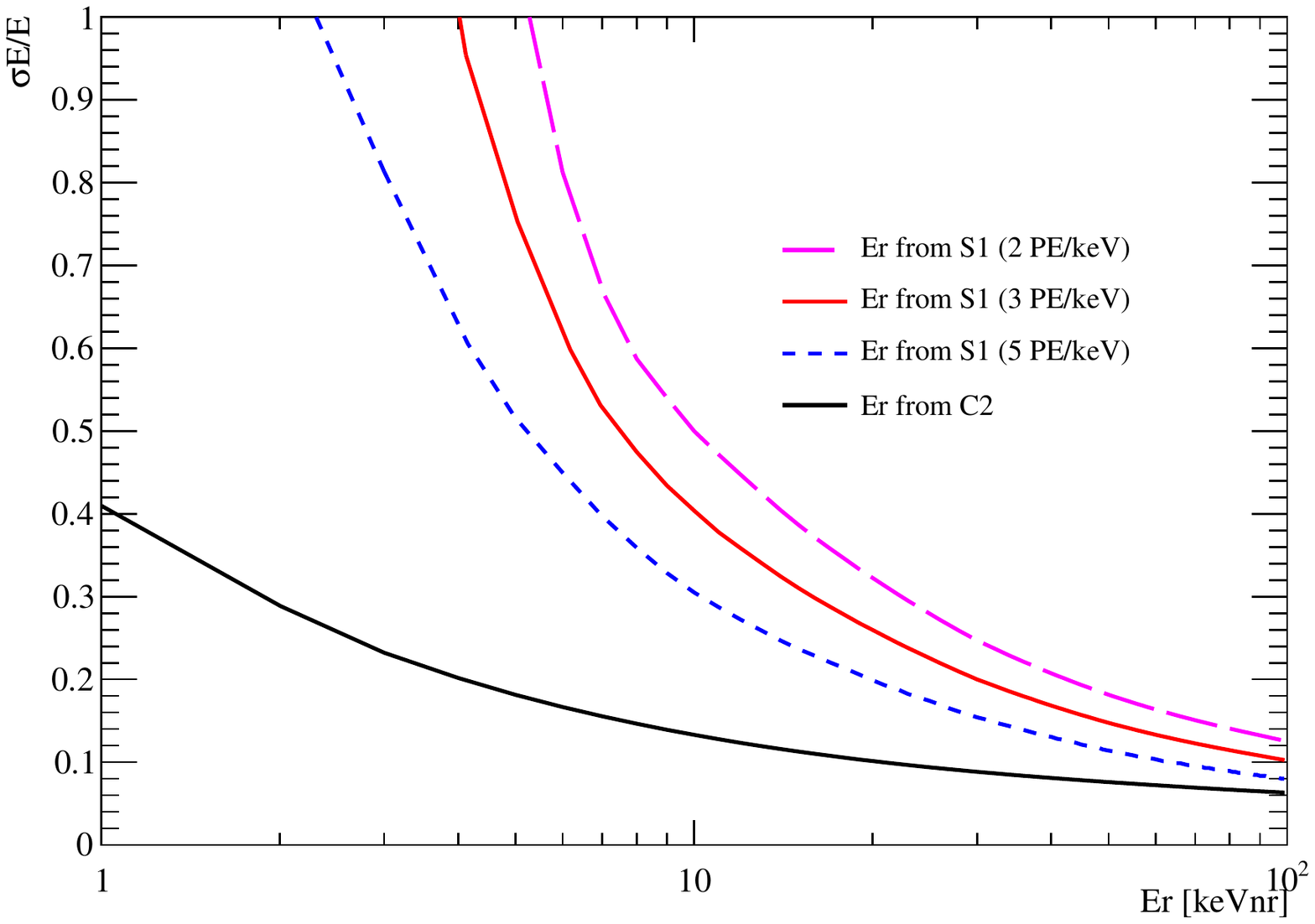}
\caption{\small{Energy resolution as a function of recoil energy for an energy scale determined using S1 (for three different light yields) and C2. Above $\sim$40~\kevnr, the resolutions become comparable but at lower recoil energies, of relevance to WIMP searches, the advantage of a C2 derived energy scale is evident.}}
\label{fig:resolution}
\end{center}
\end{figure}

Fig.~\ref{fig:s1c2energy} shows the simple relationship between S1, C2 and nuclear recoil energy. Here three different cases, \ly~=~2, 3, \& 5 PE/\kevee, are shown for S1.  For a given energy, C2 is an order of magnitude larger than S1, and the advantage of C2 over S1 grows for lower energies. Fig.~\ref{fig:resolution} shows the energy resolution ($\sigma$/E) of E1 and E2.  Here again the three cases of \ly~=~2, 3, \& 5 PE/\kevee\ are presented. Even with \ly\ as large as 5 PE/\kevee, E1 becomes severely limited below $\sim$10~\kevnr.  On the other hand, E2 shows reasonable resolution ($\sim$20\%) down to 3 \kevnr.

\section{Comparison of the different phase spaces} 
\label{sec:phase_spaces}

To visualize the problem of the conventional method, log$_{10}$(S2/S1) vs. S1, four phase spaces are contrasted:
\begin{enumerate}[label=\Roman{*}.]
\item   ~S2 vs. S1
\item   ~log$_{10}$(S2/S1) vs. S1
\item   ~log$_{10}$(S1/C2) vs. C2
\item   ~log$_{10}$(E1/E2) vs. E2
\end{enumerate}

Recall that S1 is analytically connected to E1 by \leff, while C2 is analytically connected to E2 by \qy. Thus there exists a strict one-to-one mapping between all four phase spaces listed. 

\subsection{Monoenergetic nuclear recoil injection}

As a simple illustration of these phase spaces, we first inject monoenergetic nuclear recoil energies of 4, 8, 16 and 32~\kevnr. Such energies are representative of the range of interest for WIMP scatters in LXe. Typically detectors are calibrated with broad spectrum sources such as Am-Be or $^{252}$Cf neutron emitters that produce recoils across this energy range. Note that the results are of course unaffected for the case of continuous spectra.

\begin{figure}[htb!]
\begin{center}
\includegraphics[width=.5\textwidth]{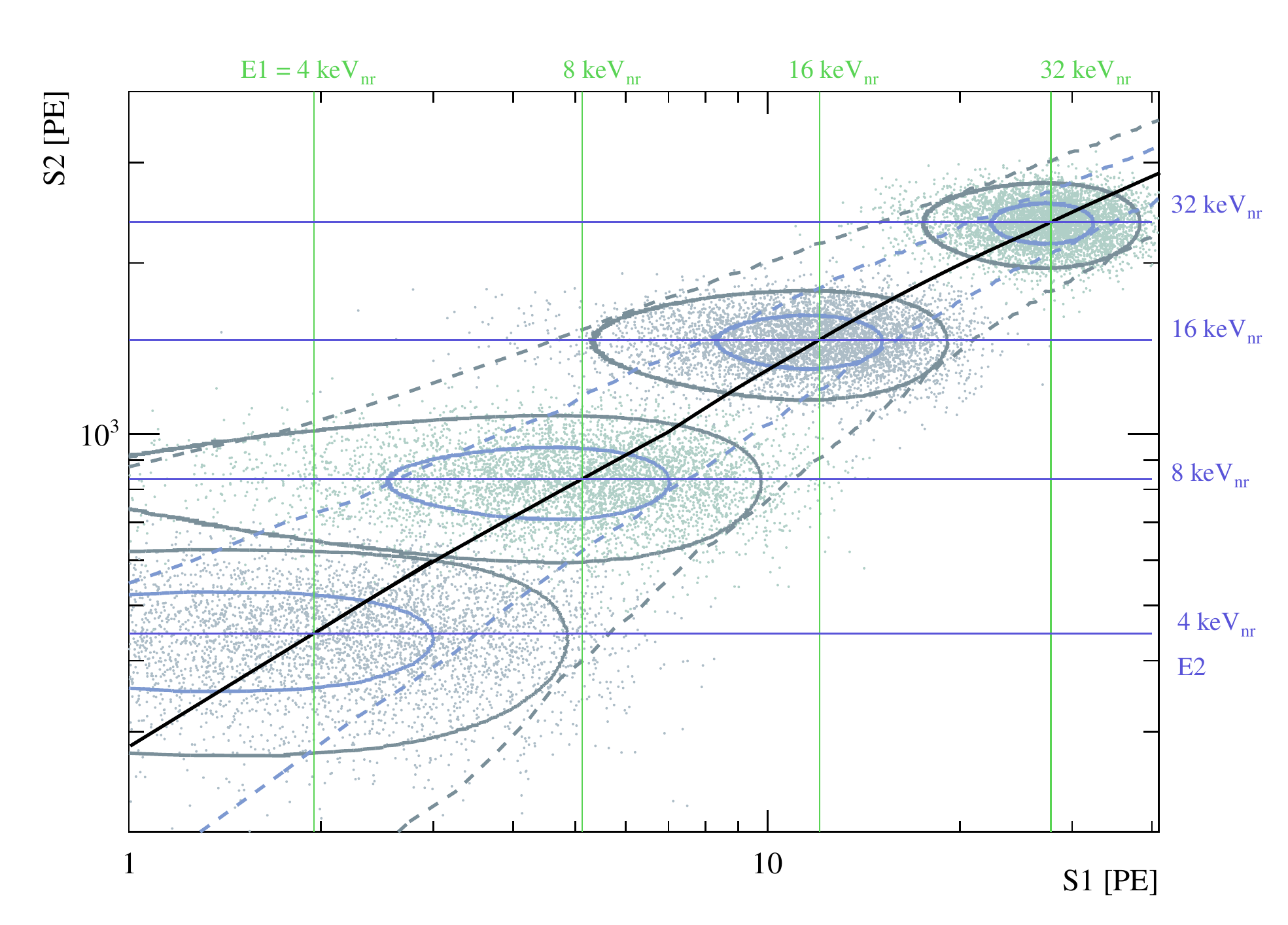}
\caption{\small{The relationship between S1 and S2 for monoenergetic nuclear recoils of 4, 8, 16, and 32~\kevnr. Solid contours represent 1$\sigma$ (blue) and 2$\sigma$ (grey) deviations from the mean for these monoenergetic neutrons. Dashed curves represent the 1$\sigma$ (blue) and 2$\sigma$ (grey) deviations from the mean from continuous energy recoils. Green lines (vertical here) indicate energy in E1 and blue lines (horizontal here) indicate energy in E2. Note that the same convention is used throughout this article. It is clear that the Poisson fluctuations in S1 gives a greater distortion for lower recoil energies making an energy scale determined by S1 significantly more ambiguous as true recoil energy gets lower.}}
\label{fig:s2s1}
\end{center}
\end{figure}

Firstly, Fig.~\ref{fig:s2s1} shows S2 vs. S1. Here, as well as throughout the article, the following are indicated:
\begin{itemize}
\item For each of the sets of monoenergetic energy recoils, 1$\sigma$ and 2$\sigma$ regions are shown as contours.   
\item A black solid curve indicates E2 = E1. Additional $\pm$1$\sigma$ and $\pm$2$\sigma$ curves show the band of continuous nuclear recoil energy distribution (as from, for {\em e.g.}, Am-Be neutron calibration data).
\item E1 = 4, 8, 16, 32~\kevnr\ are indicated with green lines, while E2 = 4, 8, 16, 32~\kevnr\ are indicated with blue lines.   
\end{itemize}

It is clearly seen that at lower energy the S1 distribution is much wider than the S2 distribution, highlighting the fact that S2 is the better energy estimator.  At 4~\kevnr, S1 is only 2 PE, thus its Poisson fluctuation prohibits any useful energy estimate. 

\begin{figure}[htb!]
\begin{center}
\includegraphics[width=.5\textwidth]{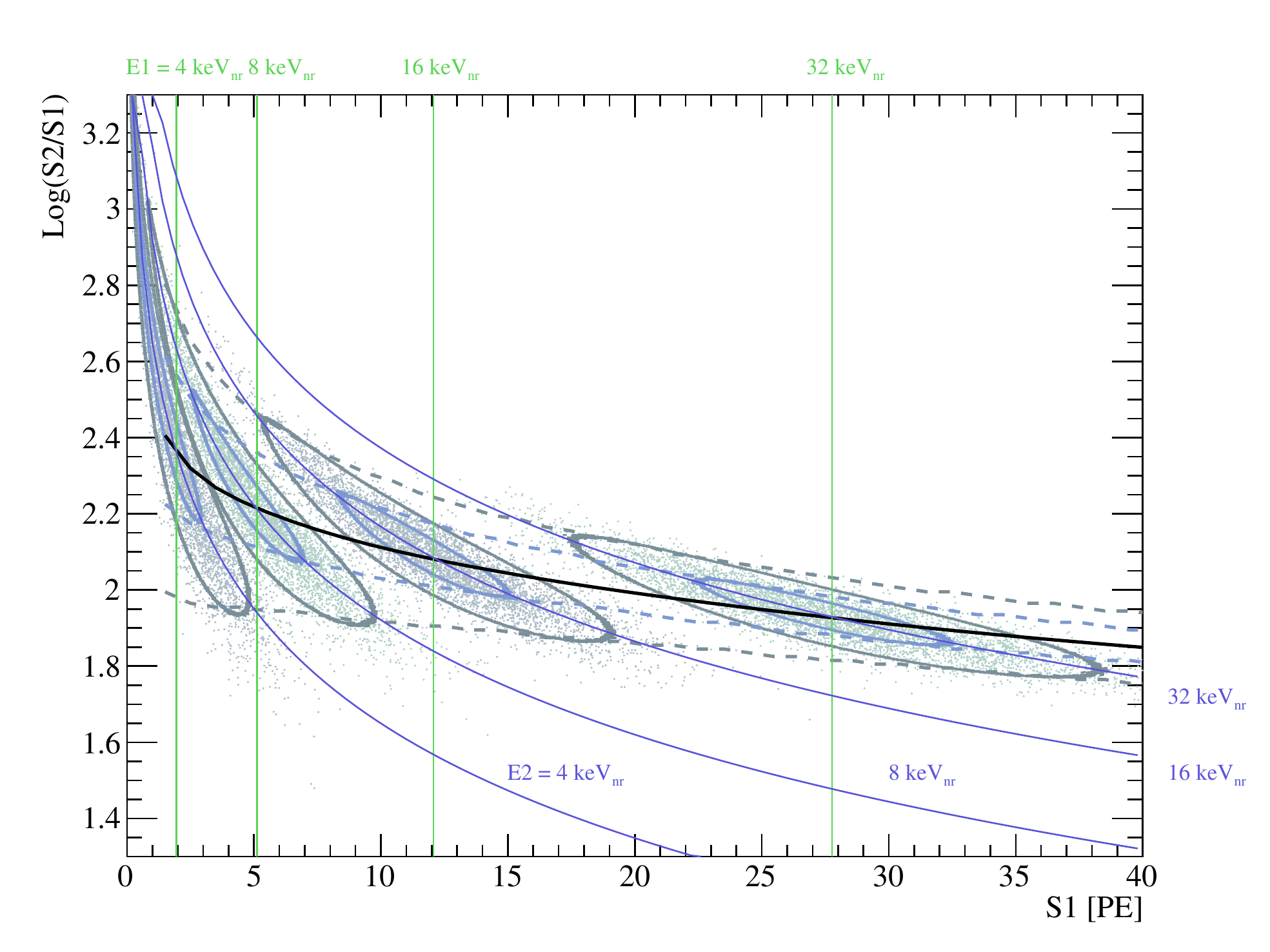}
\caption{\small{The distribution of monoenergetic nuclear recoils as shown in traditional log$_{10}$(S2/S1) vs. S1 parameter space. The distortion of what has been previously used to represent a recoil energy scale is seen. The S1 is not a well constrained representative of energy on an event-by-event basis.}}
\label{fig:s2s1vs1}
\end{center}
\end{figure}

Next, Fig.~\ref{fig:s2s1vs1} shows the traditional phase space of log$_{10}$(S2/S1) vs. S1 for the monoenergetic recoils.  It is seen that the recoils are distributed in an elongated ellipse for any given energy. The ellipses are aligned with contours of constant E2 because E2 represents the nuclear recoil energy with less spread (and ambiguity) than S1. It is evident that this phase space is distorted, and thus introduces a corresponding distortion to the value energy calculated by a measurement of S1.

Fig. 8 introduces the new phase space of log$_{10}$(S1/C2) vs. C2.  Unlike in the log$_{10}$(S2/S1) vs. S1 phase space (Fig.~\ref{fig:s2s1vs1}), each energy band is now well aligned as a vertical ellipse. Note that they are still well separated, as in Fig.~\ref{fig:s2s1} showing S2 vs. S1. Note also the simple relation log$_{10}$(S1/C2) = log$_{10}$(S1/(S2/20)) = 1.30 - log$_{10}$(S2/S1). Therefore the y-axis of this new phase space, log$_{10}$(S1/C2), is merely a mathematical transformation of log$_{10}$(S2/S1). The critical advantage of this method comes from adopting the new x-axis of C2 instead S1, whilst retaining discrimination.

\begin{figure}[htb!]
\begin{center}
\includegraphics[width=.5\textwidth]{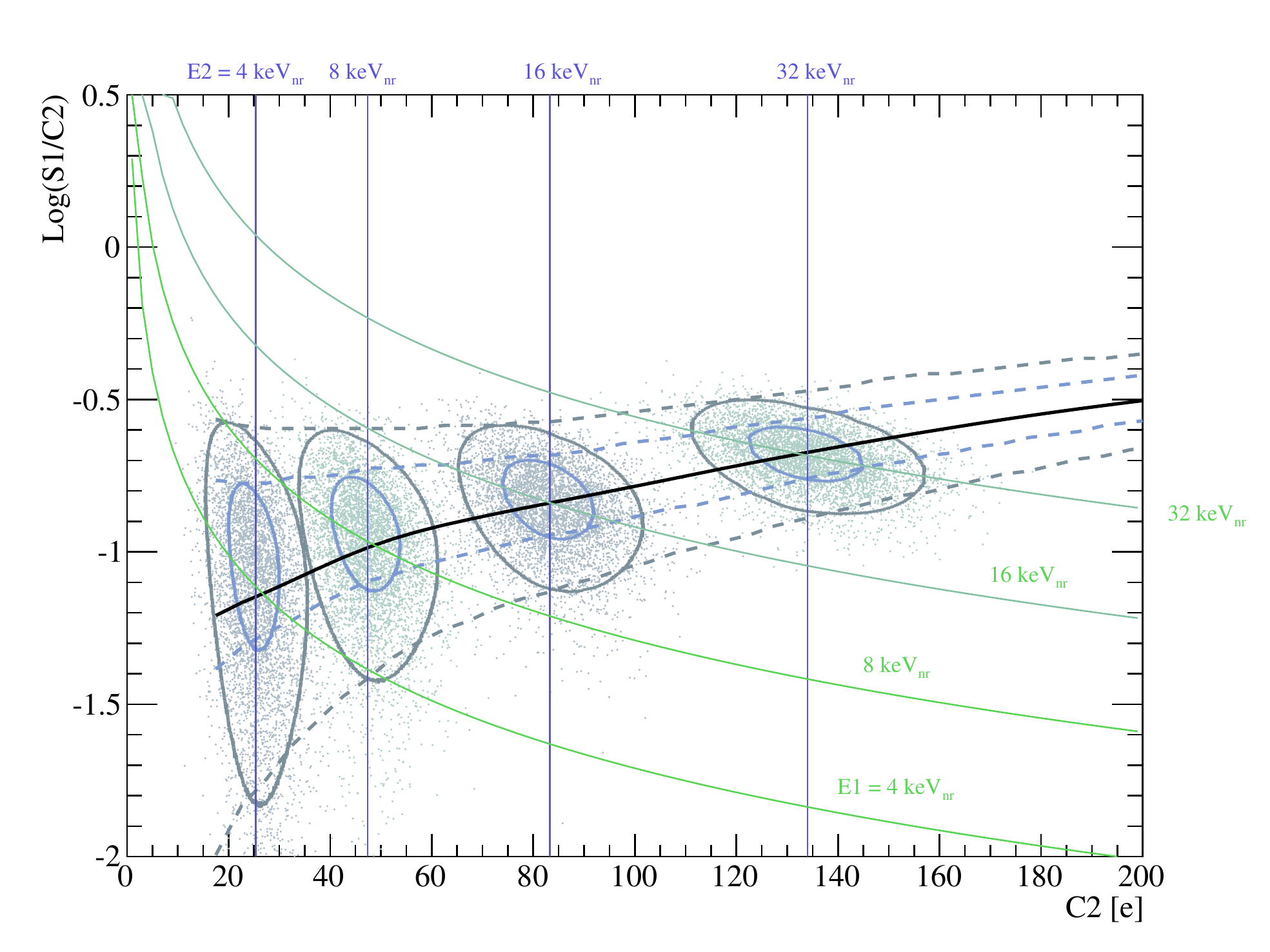}
\label{fig:s1c2vc2}
\caption{\small{The monoenergetic nuclear recoils represented in the log$_{10}$(S1/C2) vs. C2 phase space. Transposed onto the C2 scale, the ambiguity of the conversion between an observable and a recoil energy is significantly reduced, with no loss of discrimination.}}
\end{center}
\end{figure}

\begin{figure}[htb!]
\begin{center}
\includegraphics[width=.5\textwidth]{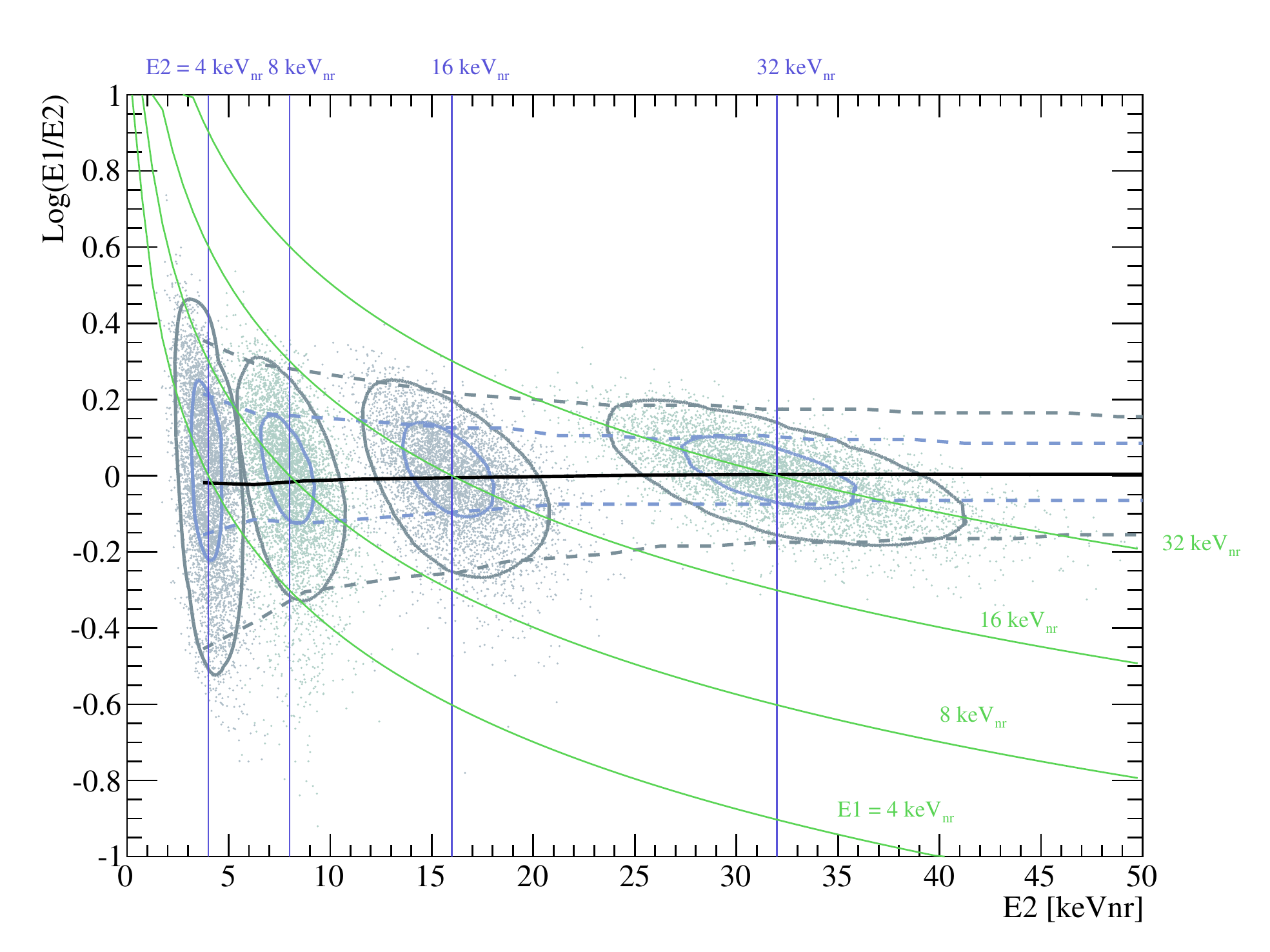}
\caption{\small{For the comparison of observed and expected WIMP recoil signals, it is necessary to translate into a true nuclear recoil energy scale, as represented here in log$_{10}$(E1/E2) vs. E2 phase space.}}
\label{fig:e1e2ve2}
\end{center}
\end{figure}

\begin{figure}[htb!]
\begin{center}
\includegraphics[width=.5\textwidth]{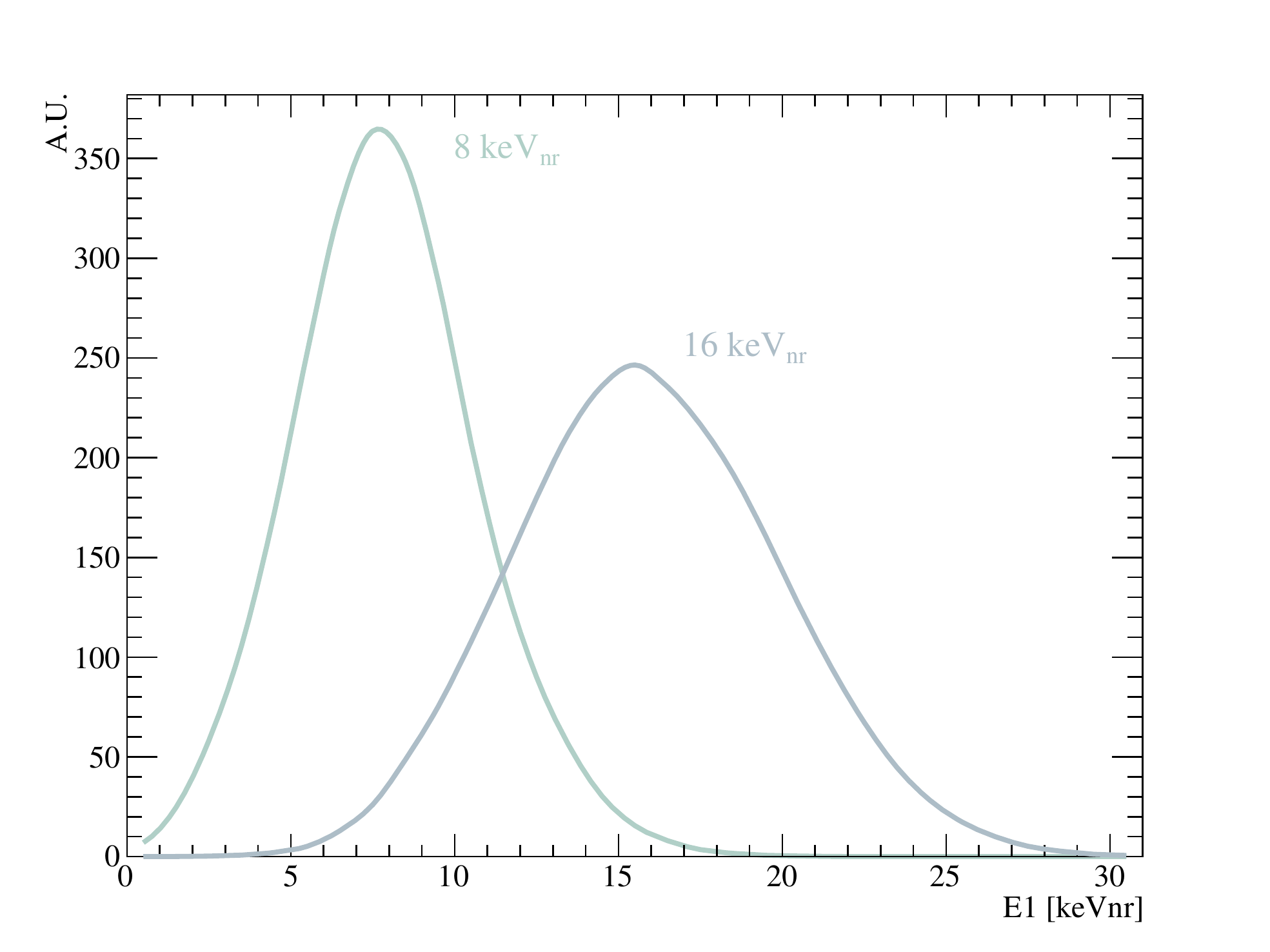}
\includegraphics[width=.5\textwidth]{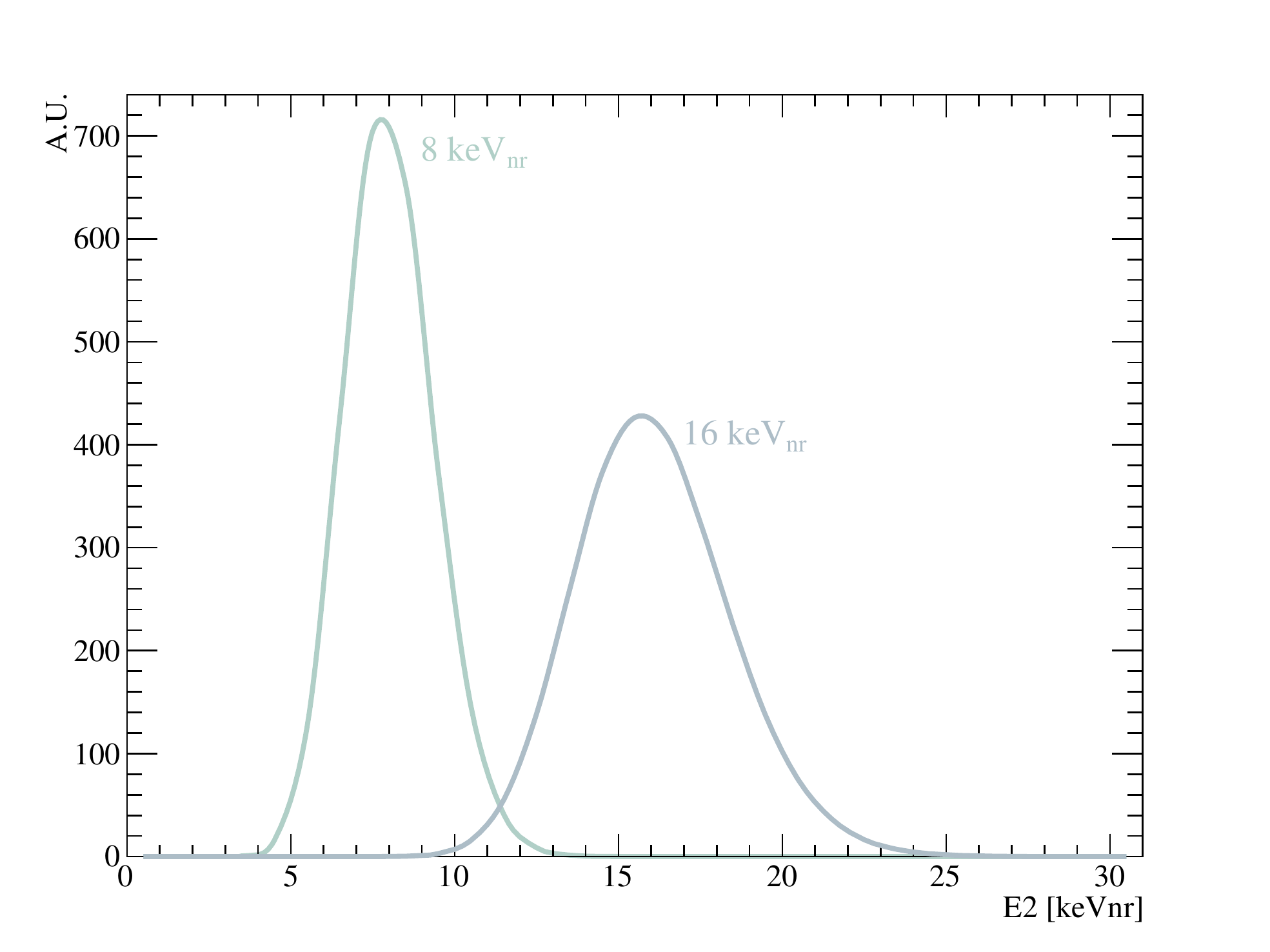}
\caption{\small{ Comparison of 8~\kevnr\ and 16~\kevnr\ nuclear recoil energy spectra in E1 (top) and E2 (bottom), highlighting the superior energy resolution in E2 that improves relative to E1 for falling energies.}}
\label{fig:e1e2}
\end{center}
\end{figure}

Finally, in Fig.~\ref{fig:e1e2ve2}, we introduce the phase space of log$_{10}$(E1/E2) vs. E2:

\begin{itemize}
\item The x-axis directly represents the best estimate of the nuclear recoil energy on an event-by-event basis.  Fig.~\ref{fig:e1e2} depicts the superior energy resolution of the E2 phase space.
\item log$_{10}$(E1/E2) = 0 ({\em i.e.}, E2~=~E1) sets the mean for all the nuclear recoil events, in the absence of imposed thresholds.  
\item Experimental specific calibrations are incorporated into the only physically meaningful parameters, E2 (energy) and E1/E2 (discrimination); this phase space is not specific to any single detector.
\end{itemize}

\subsection{Representing the traditional signal box}

To further evaluate the benefits and disadvantages of the four different phase spaces, let us now consider a `signal box' for a WIMP search. Traditionally the WIMP signal box is defined by: 
\begin{itemize}
\item The accepted energy range, determined by a minimum and maximum energy of E1 (given by S1).
\item The neutron band given by Am-Be calibration data. In log$_{10}$(S2/S1), events below the mean of the distribution (where discrimination is highly efficient) and within 2$\sigma$ or 3$\sigma$ are accepted.
\end{itemize}

As a typical and realistic example, let us assume: 
\begin{itemize}
\item An E1 range of 8 -- 32~\kevnr\
\item The neutron population in the band between the mean and 2$\sigma$ in log$_{10}$(S2/S1) vs. S1.
\end{itemize}

\begin{figure*}[htb!]
\begin{center}
\includegraphics[width=.49\textwidth]{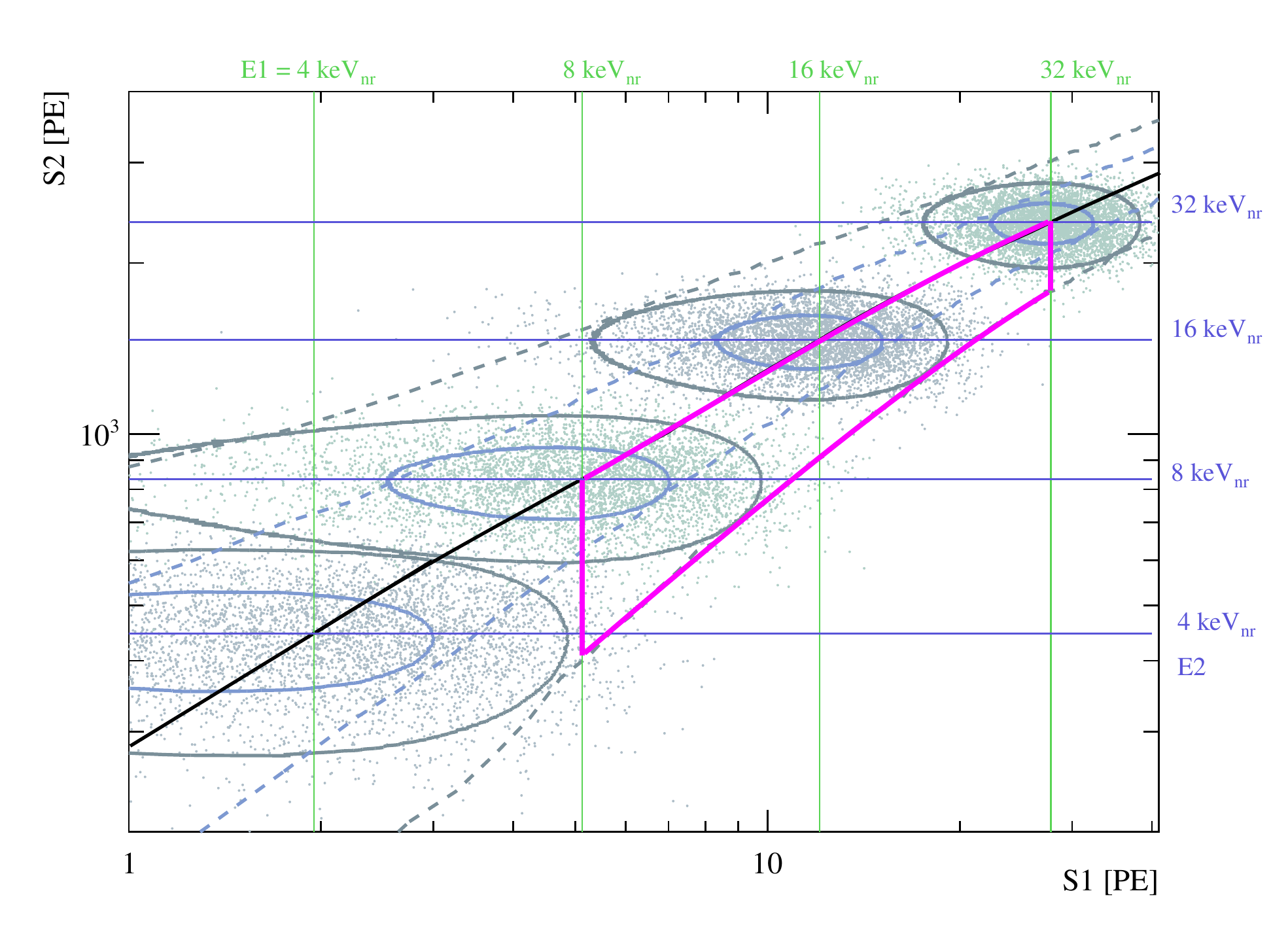}
\includegraphics[width=.49\textwidth]{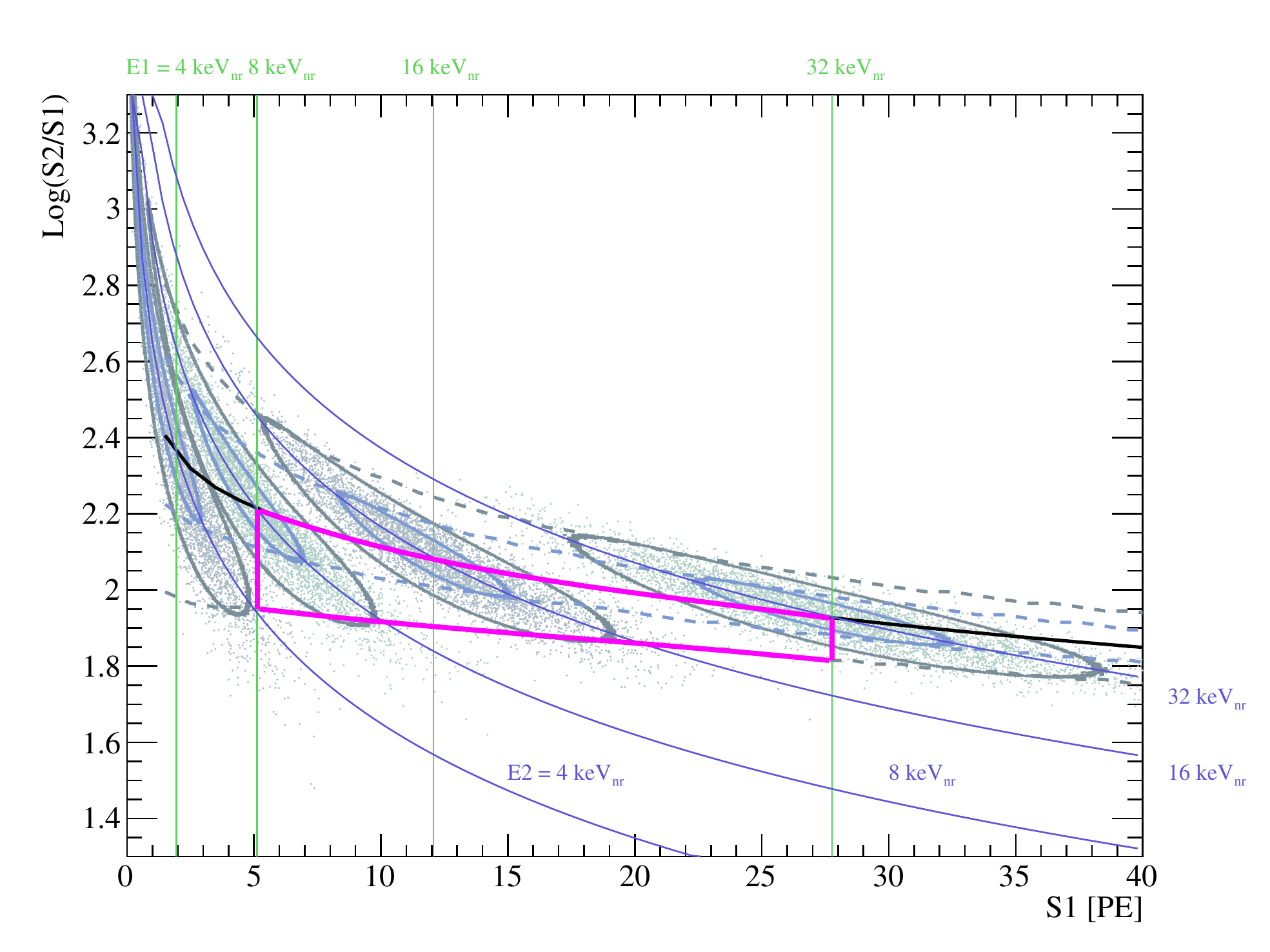}
\includegraphics[width=.49\textwidth]{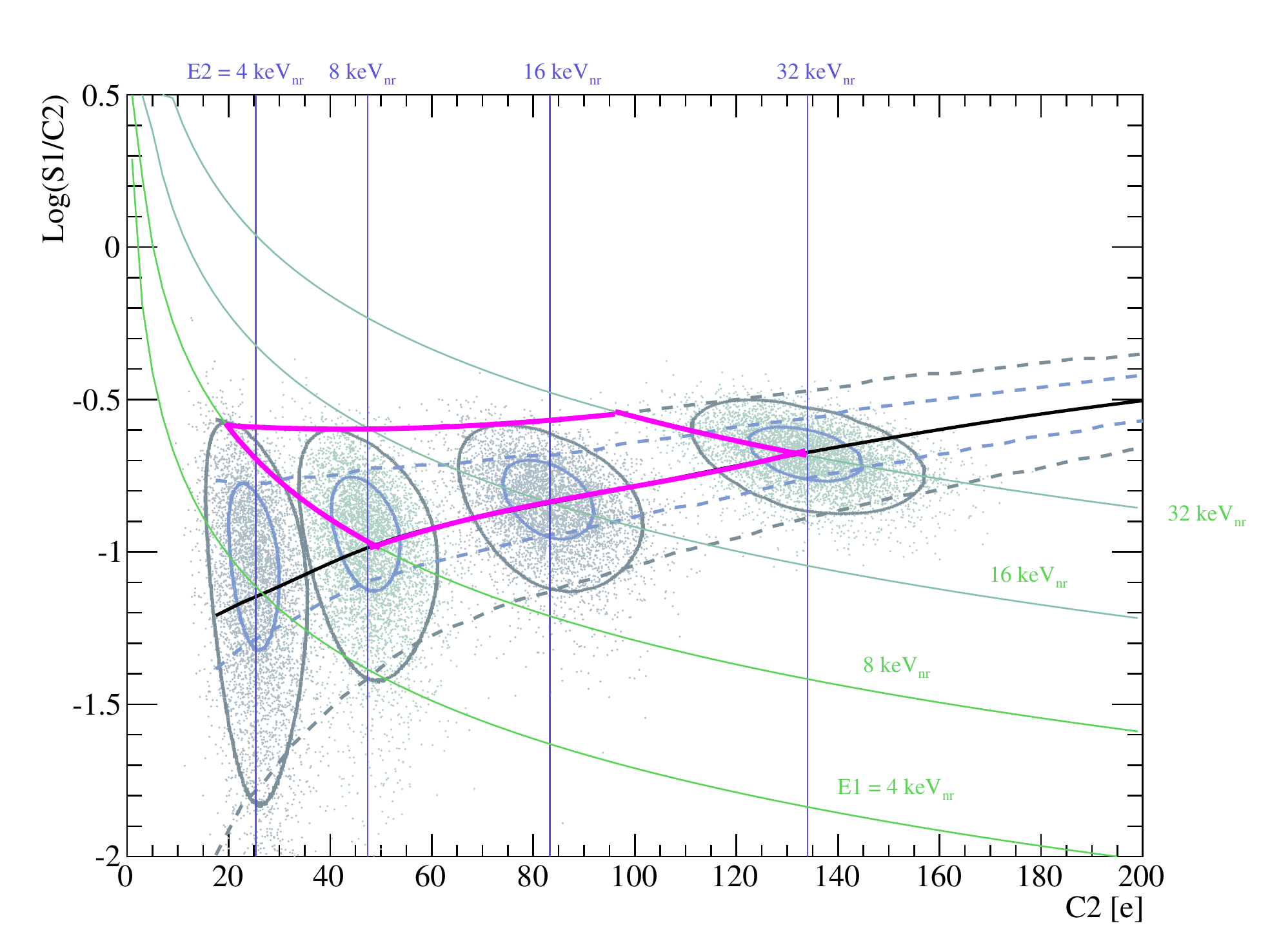}
\includegraphics[width=.49\textwidth]{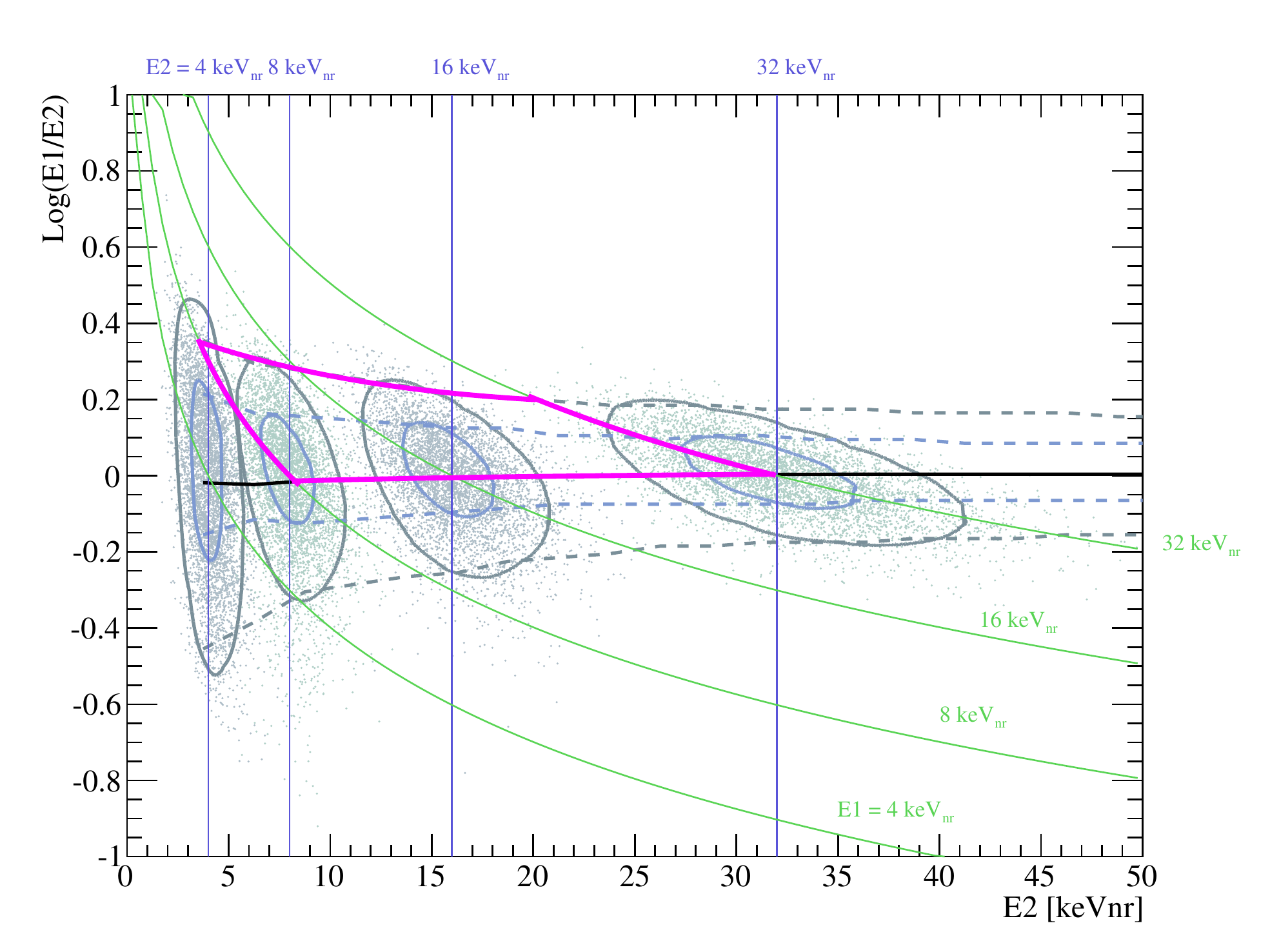}
\caption{\small{The traditional signal box (pink boundaries) as defined in log$_{10}$(S2/S1) vs. S1 and applied to the four phase spaces described in the text. It can be seen that WIMP recoil acceptance at 4 and 32~\kevnr\ is severely reduced in relation to that suggested in the S1 space once the box is transposed into the C2 or E2 energy scales.}}
\label{fig:traditional_box}
\end{center}
\end{figure*}

This traditional signal box is shown in Fig.~\ref{fig:traditional_box} as contained within the pink borders in all four phase spaces. There are two important remarks to be made:
\begin{enumerate}
\item The maximum energy of 32~\kevnr\ is indicated in E1 as a vertical line in the traditional log$_{10}$(S2/S1) vs. S1 phase space.  However, only a very small fraction of 32~\kevnr\ events are actually accepted by the signal box (see Fig.~\ref{fig:acceptance}).
\item At the minimum energy of 8~\kevnr, the bound is rather loose towards low energy events.  There are some fraction of events between 4 and 8~\kevnr\ that penetrate into the signal box. This is the origin of the so-called `Poisson smearing' that enhances the sensitivity of such detectors to low energy events, and thus to low mass ($<$10 GeV/c$^{2}$) WIMPs whilst maintaining a relatively high threshold (8~\kevnr\ in this model). Here some fraction of events that deposit energies below threshold may be detected in the signal region following statistical fluctuations in the number of detected PE.
\end{enumerate}

It is worth noting that the traditional signal region shows up as quite a distorted box in both log$_{10}$(S1/C2) vs. C2 and log$_{10}$(E1/E2) vs. E2, as shown in Fig.~\ref{fig:traditional_box}. 

\subsection{Introduction of a new signal box}

\begin{figure*}[htb!]
\begin{center}
\includegraphics[width=.49\textwidth]{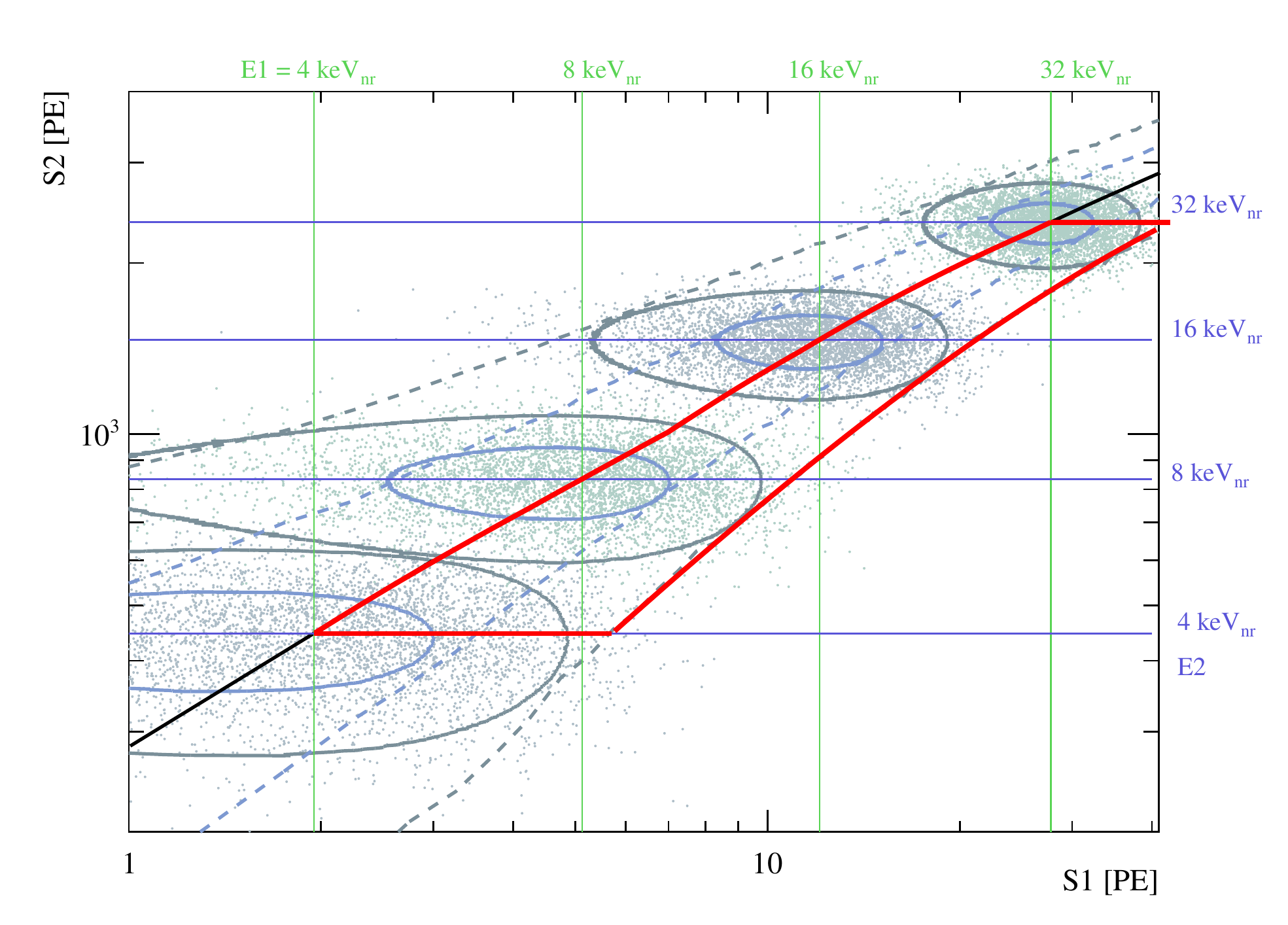}
\includegraphics[width=.49\textwidth]{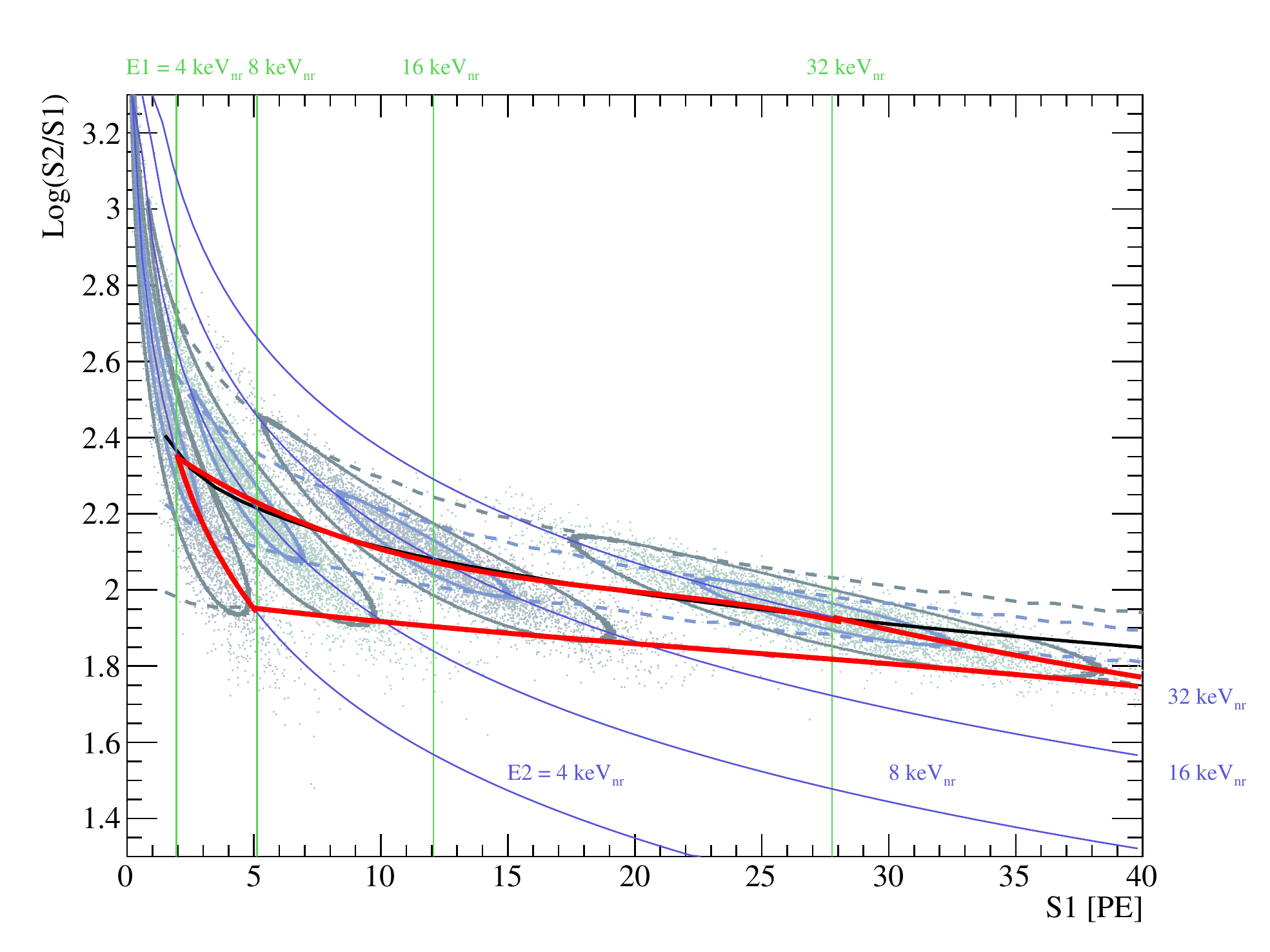}
\includegraphics[width=.49\textwidth]{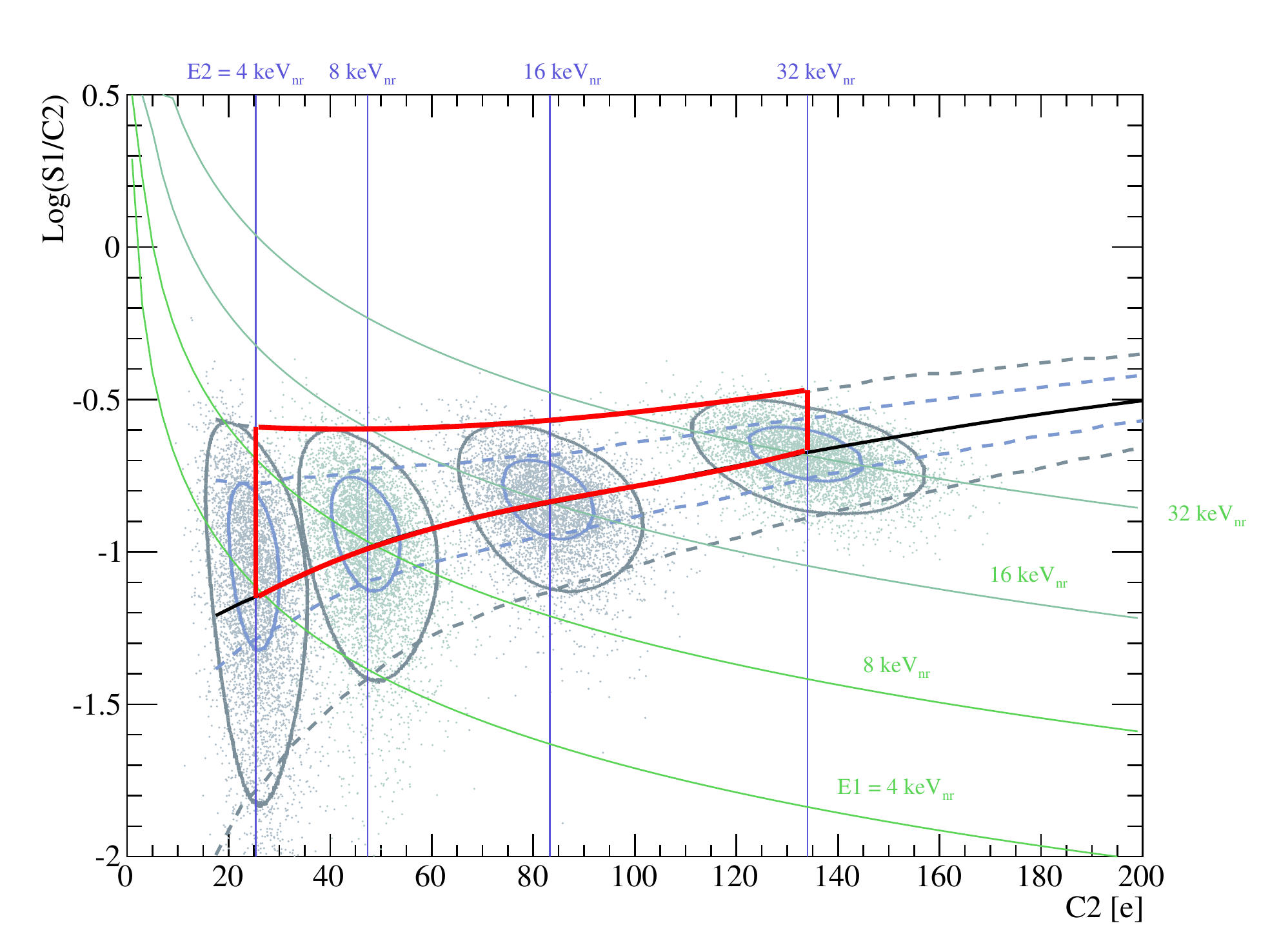}
\includegraphics[width=.49\textwidth]{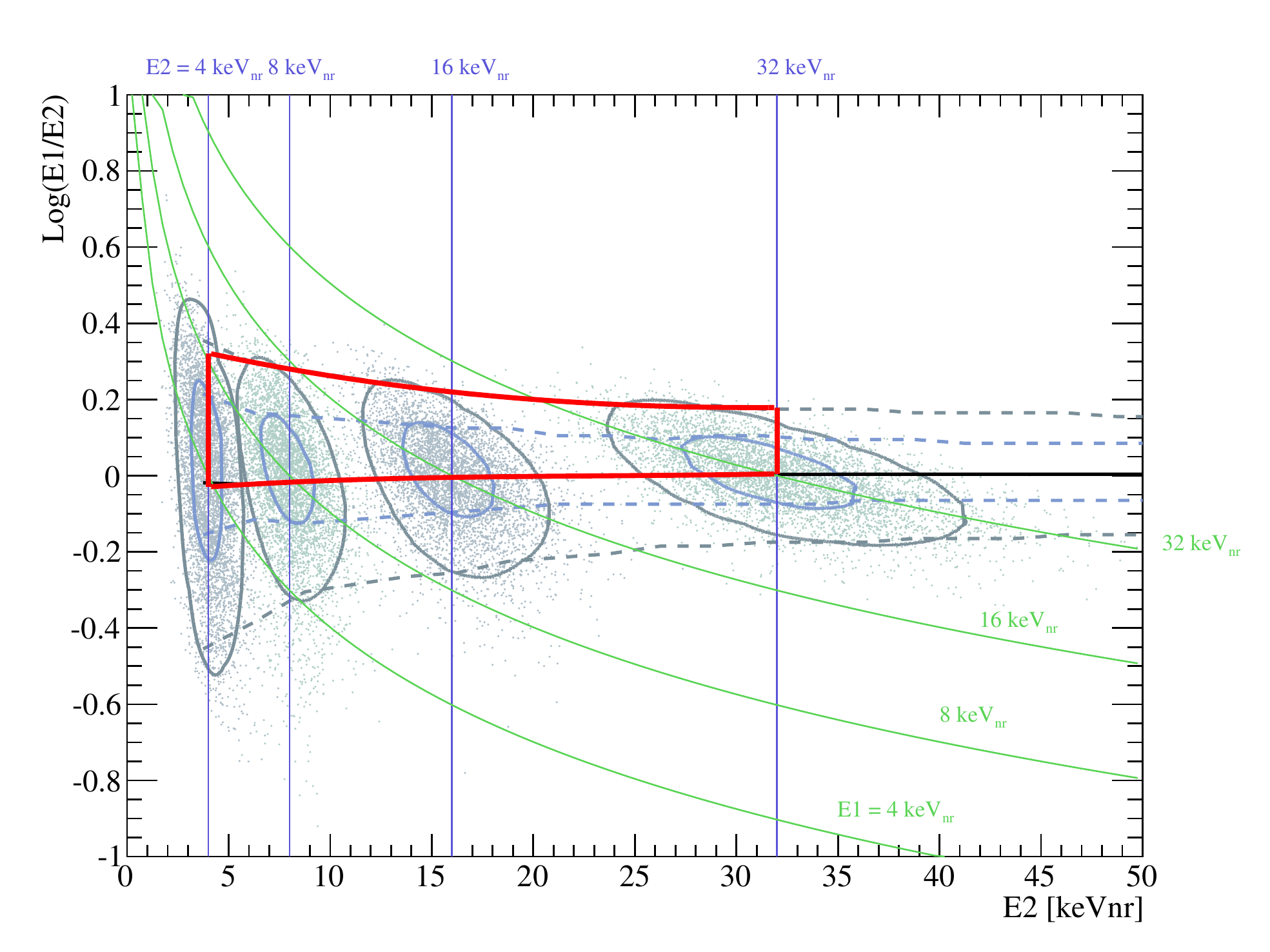}
\caption{\small{A new signal box (red boundaries) is defined in log$_{10}$(E1/E2) v E2 phase space and applied to the four phase spaces described in the text. Applying this box into traditional phase space of log$_{10}$(S2/S1) vs. S1 serves to highlight the huge distortion present in the traditional method of signal box definition and analysis. The C2 and E2 energy scales also allow measurements of discrimination parameters and estimation of background leakage into the signal region from slices in C2 specific to a given energy.}}
\label{fig:new_box}
\end{center}
\end{figure*}

To address the fact that E1 is not a good energy estimator, a new signal box, based on E2, is defined. In the case of E2, the energy can be well determined even down to the 2 -- 3~\kevnr\ range.  Therefore, there is no reason to define a lower bound at 8~\kevnr\ (which is necessary in the case of the E1-based phase space to ensure good signal acceptance at threshold).  Instead, a minimum energy of 4~\kevnr\ (close to the lowest directly measured relative scintillation and ionization yield values) is introduced in this model.  The signal box that is proposed is shown in Fig.~\ref{fig:new_box} as a red box in all four phase spaces and is defined by: 

\begin{itemize}
\item E2 range between 4 and 32~\kevnr.
\item The region between the mean and 2$\sigma$ of the neutron band in log$_{10}$(S1/C2) vs. C2 or log$_{10}$(E1/E2) vs. E2.
\end{itemize}

\begin{figure}[htb!]
\begin{center}
\includegraphics[width=.5\textwidth]{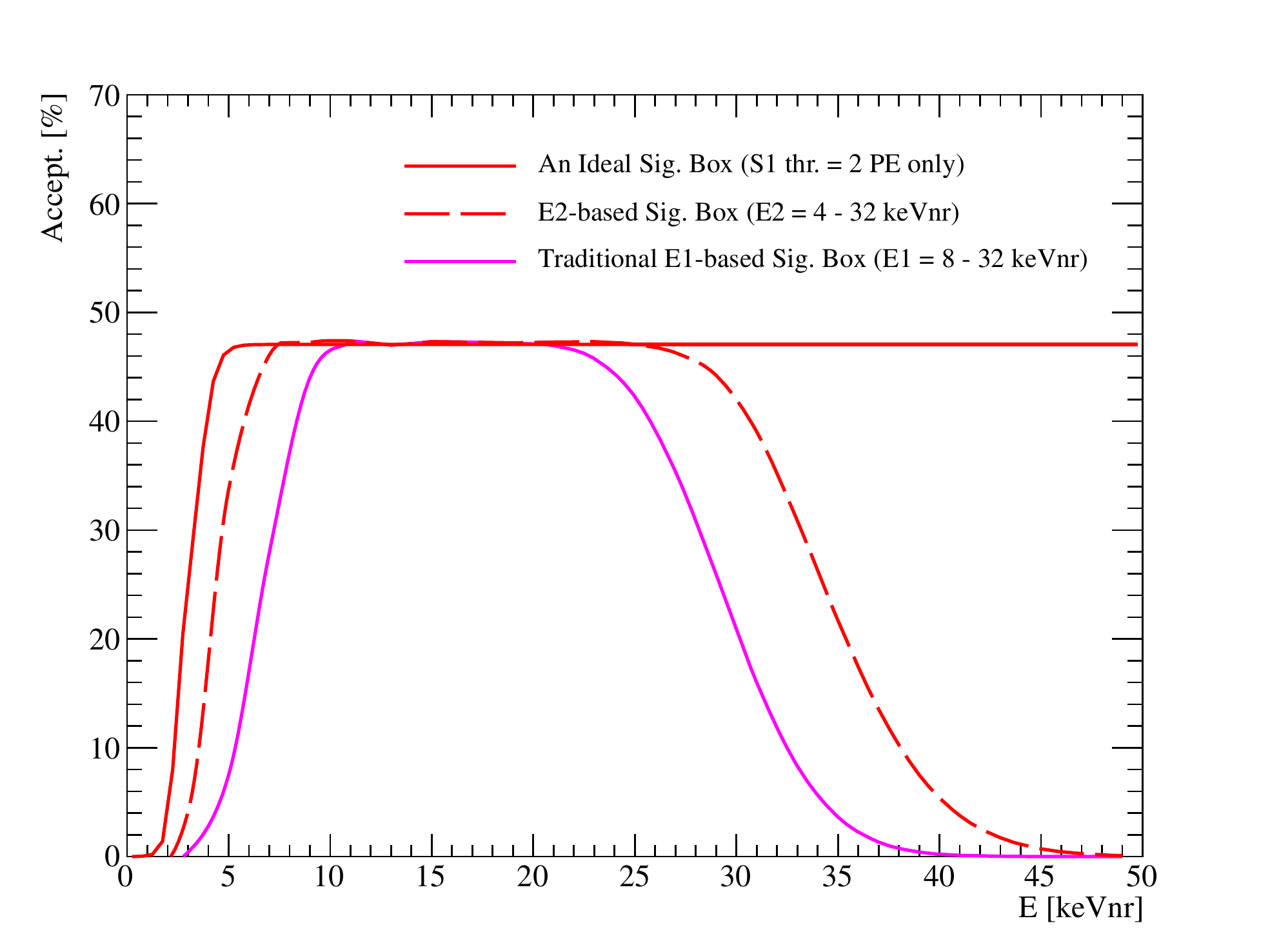}
\caption{\small{Signal acceptance in case of the traditional and newly proposed signal boxes. The E1 (S1) based acceptance (solid purple curve) when defined as 8 -- 32~\kevnr\ in S1 is inefficient at higher energies when compared to the same upper bound as defined in E2 (C2) (dashed red curve). 
The lower threshold afforded this signal box increases the efficiency at lower energies. 
Indeed, contingent upon the behavior of electromagnetic background, in principle the lower bound may be completely removed and the threshold defined simply as the minimum requirement of 2 PE ({\em i.e.}, 1 PE in 2 PMTs). No upper bound need be enforced at all.  This is represented by the solid red curve.}}
\label{fig:acceptance}
\end{center}
\end{figure}

To further illustrate the effect of minimum and maximum energy bounds in the old and new methods, the signal acceptance is shown as a function of true recoil energy in Fig.~\ref{fig:acceptance}. The new signal box shows a near flat acceptance in energy from 4 -- 32~\kevnr, whilst an E1-based signal region compromises efficiency both at low and high energies.

\begin{figure}[htb!]
\begin{center}
\includegraphics[width=.5\textwidth]{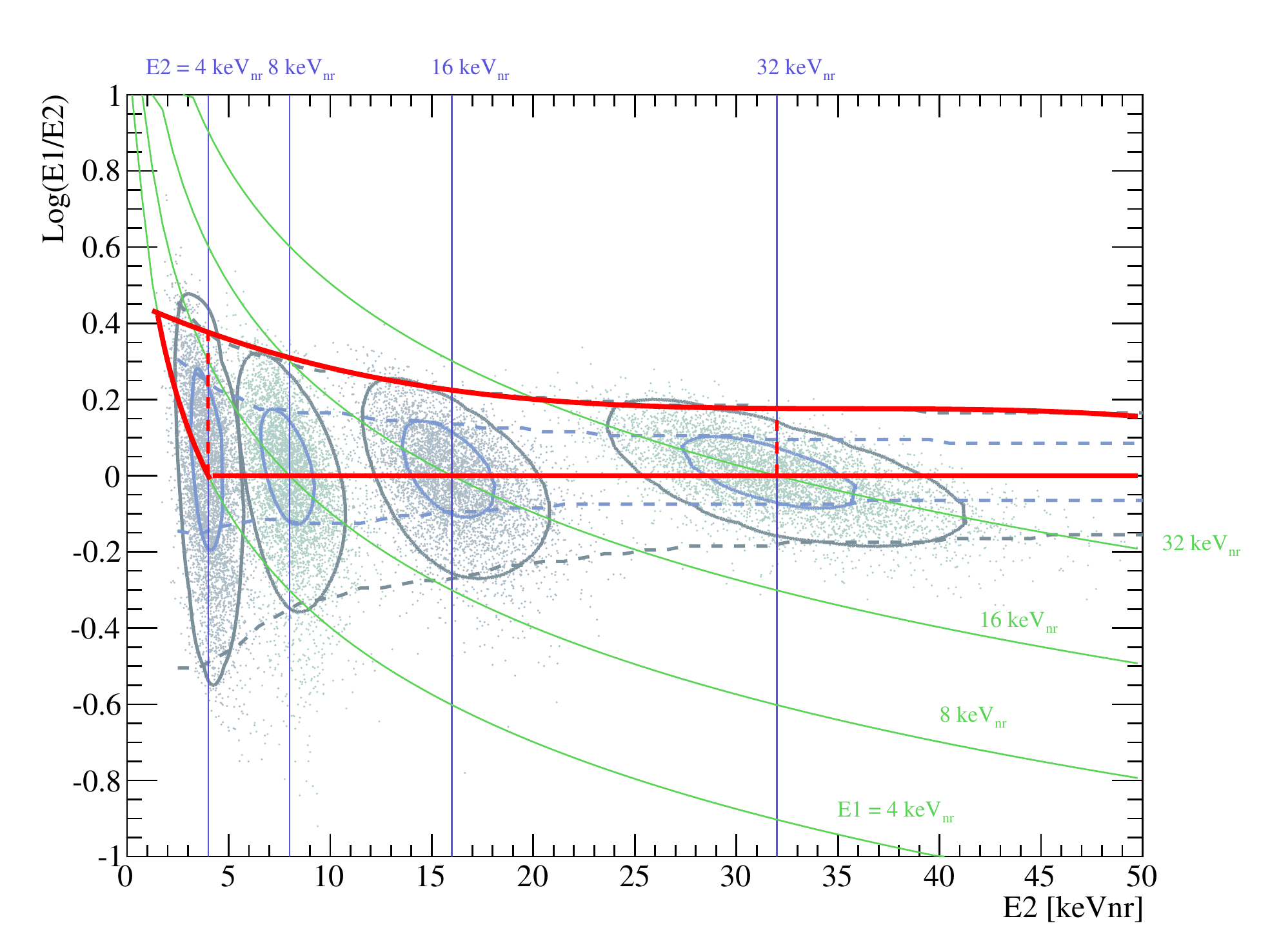}
\caption{\small{The upper and lower energy bounds in E2 that are present in Fig.~\ref{fig:new_box} are shown as red dashed vertical lines. These may be relaxed such that the signal region is defined by the solid red boundaries. Here no upper boundary is imposed, and the lower boundary is defined by the S1 = 2 PE coincidence requirement. Acceptance to WIMP recoils is thus enhanced. This signal acceptance for this box is the corresponding solid red curve of Fig.~\ref{fig:acceptance}.}}
\label{fig:ultimate_box}
\end{center}
\end{figure}

As a final extension to the issue of signal box definition, it should be noted that in the phase space of log$_{10}$(E1/E2) vs. E2, a sharp lower energy bound at 4~\kevnr\ is unnecessary. Assuming extrapolation of models for relative scintillation and ionization yields to lower energies, or in light of future direct measurements elucidating this region, a lower bound that follows the minimum contour of 2 PE in the S1 channel may be adopted.  Similarly, there is no reason to set a higher energy bound. Indeed, a profile likelihood analysis across the 2D phase space of log$_{10}$(E1/E2) vs. E2 allows improved background rejection and signal excess energy distribution analysis as a result of the improved resolution. Such a maximally efficient signal region is presented in Fig.~\ref{fig:ultimate_box}.

\subsection{WIMP spectra injection}

\begin{figure}[htb!]
\begin{center}
\includegraphics[width=.5\textwidth]{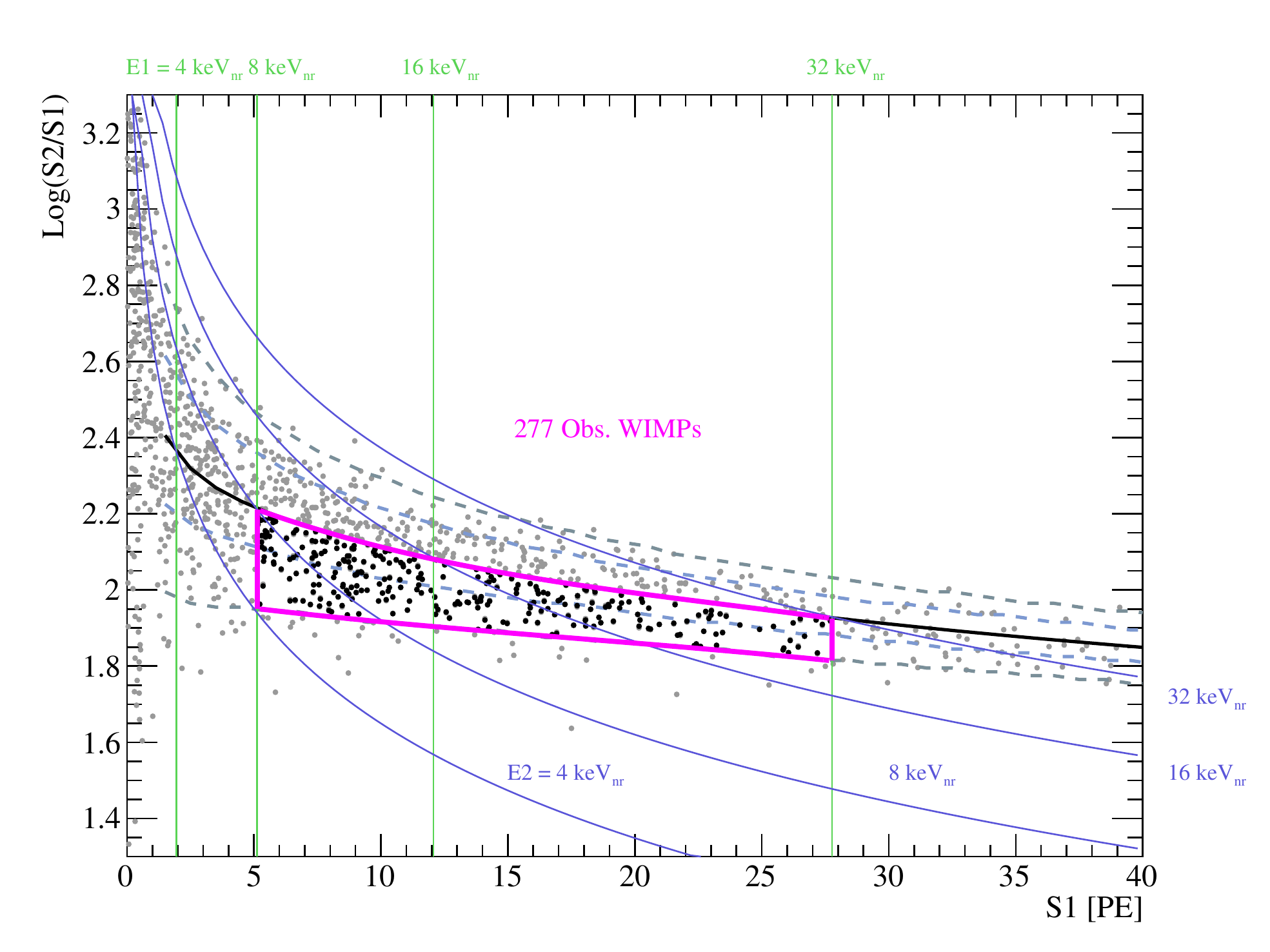}
\caption{\small{WIMP distribution in traditional phase space of log$_{10}$(S2/S1) vs. S1 (assuming 100 GeV/c$^{2}$ WIMP mass, WIMP-nucleon interaction cross-section of 5$\times$10$^{-45}$~cm$^{2}$) for a 1 ton fiducial volume and 1 year exposure. A total of 277 events are observed in the defined signal region.}}
\label{fig:wimp_traditional}
\end{center}
\end{figure}

\begin{figure}[htb!]
\begin{center}
\includegraphics[width=.5\textwidth]{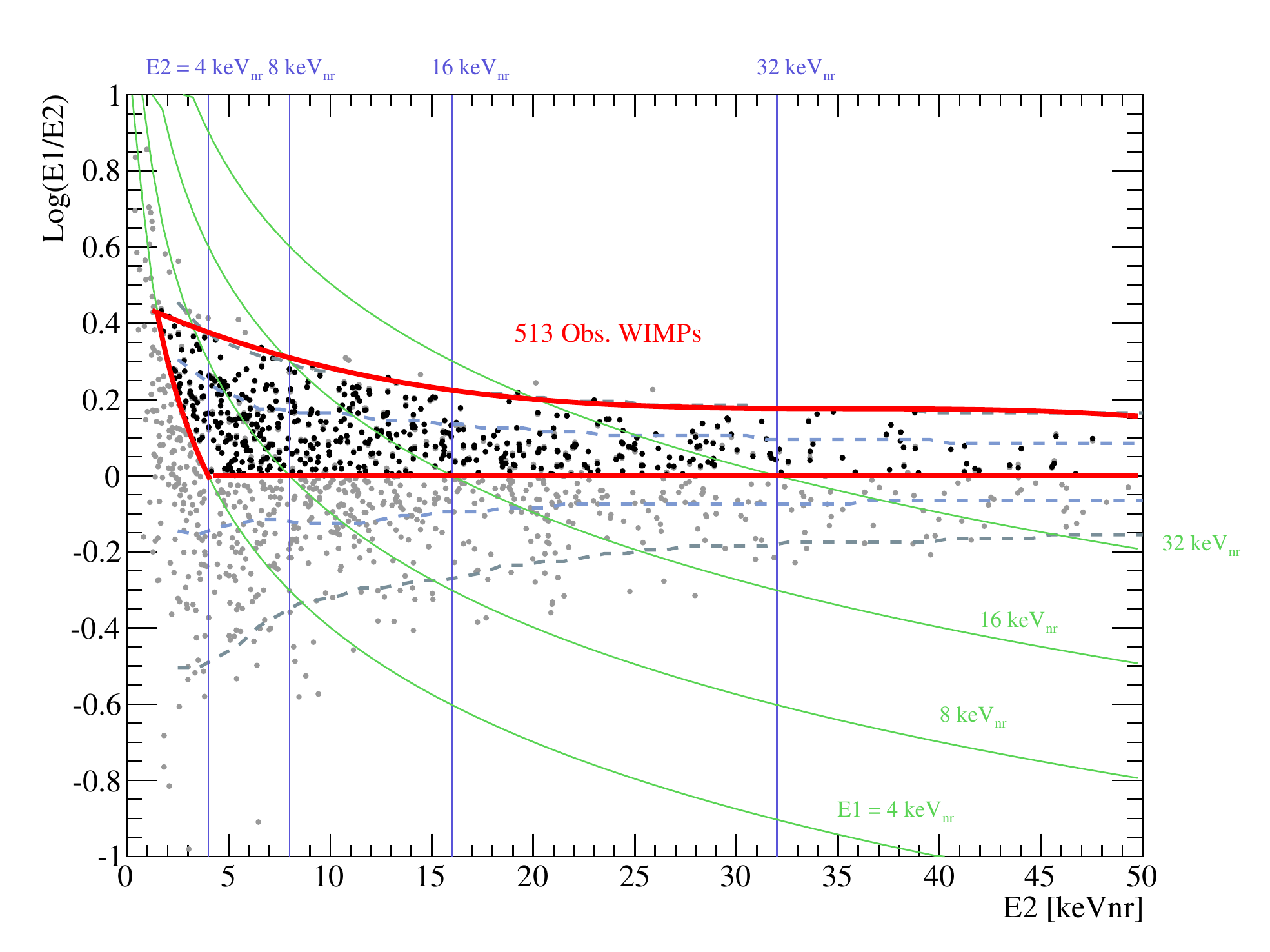}
\caption{\small{WIMP distribution in the proposed phase space of log$_{10}$(E1/E2) vs. E2 (assuming 100 GeV/c$^{2}$ WIMP mass, WIMP-nucleon interaction cross-section of 5$\times$10$^{-45}$~cm$^{2}$) for a 1 ton fiducial volume and 1 year exposure. A total of 513 events are observed in the defined signal region.}}
\label{fig:wimp_new}
\end{center}
\end{figure}

As the final stage of this investigation, injection of WIMP energy spectra are simulated. For illustration purposes a WIMP mass of 100~GeV/c$^{2}$ is assumed with a spin-independent WIMP-nucleon interaction cross-section of 5$\times$10$^{-45}$~cm$^{2}$. An exposure of 1 year for a 1 ton fiducial mass is considered.  Fig.~\ref{fig:wimp_traditional} shows the WIMP recoil event distributions in the traditional log$_{10}$(S2/S1) vs. S1 phase space and signal box as previously defined in the S1 energy estimator scale (as in Fig.~\ref{fig:traditional_box}).  Fig.~\ref{fig:wimp_new} shows the same WIMP event distribution in log$_{10}$(E1/E2) vs. E2 with the signal box as defined with a lower threshold dictated only by the 2 PE requirement for S1 (as in Fig.~\ref{fig:ultimate_box}). The lower energy threshold from the E2 scale and signal box bounded only by the S1 threshold naturally results in significantly greater signal acceptance for WIMPs, whilst maintaining discrimination: of order 500 events are observed as compared to approximately 300 for the traditional box. 

Furthermore, with the E2 energy scale, given appropriate statistics, the new representation may expose an exponential energy spectrum with a WIMP mass-dependent slope with greater confidence than the traditional S1 energy scale. This has an important consequence on the discovery of WIMPs. The new method of log$_{10}$(E1/E2) vs. E2 allows investigation of not only the excess of events over expected background, but also the energy spectrum of these events and determination of WIMP mass.

\section{Discussion and Conclusions}

\subsection{Discussion}

The new concept of log$_{10}$(E1/E2) vs. E2 as a replacement of the current standard of log$_{10}$(S2/S1) vs. S1 has been introduced based on Monte Carlo simulations of a theoretically motivated and detector realistic model.  However, the true power of the new analysis method can be realized only in conjunction with real data analysis. In particular, the method makes for a considerably more stringent probe of accurate reproduction of neutron calibration data by Monte Carlo simulations. Demanding coherence across all four types of phase space, systematic comparison between data and Monte Carlo simulations will allow considerably more robust estimates of nuclear recoil acceptance, as well as verification of assumptions in the energy dependence of \leff\ and \qy.

One should note, however, that the most reliable tests of such models could be performed by direct calibration of dark matter detectors {\em in situ}, using monoenergetic neutron generators, very much in the same manner as dedicated \leff\ measurements. 

With well modeled neutron calibration data, reproduced by simulation across all phase spaces, the detection efficiency can be reliably measured down to very low energy. Provided background remains negligible, the lowest possible energy threshold in both S1 and C2 can be adopted.  Although for C2 this could be as low as several electrons ($\sim$\kevnr), assuming sufficient electron mobility in the liquid, the minimum requirement of S1 = 2 PE nonetheless dominates. 

The proposed approach also has consequence for background expectation analysis. In reality, since the energy may be determined more reliably with the new method, the background distribution in the new phase space (either log$_{10}$(S1/C2) vs. C2 or log$_{10}$(E1/E2) vs. E2) is more meaningful.  The traditional approach of estimating background in slices of fixed S1 is misleading because in each such slice, the energy of events is not constant; the larger the value of (S2/S1), the larger the energy. Furthermore, the new phase space may be of considerable benefit in characterising rare topology background, such as so-called Multi-Scintillation Single-Ionization (MSSI) events. These are predominantly $\gamma$-ray multiple scatter events where one vertex is in the bulk volume, producing both scintillation (S1) and ionization (S2), but the second is in a region with no applied drift field (such as below the TPC cathode). Partial scintillation (S1', E1') may be detected from such `inactive' volumes whereas no electrons are extracted into the gas phase meaning no ionization signal. Such events appear with artificially large S1 signals (from both interaction sites) relative to the ionization signal (from a single site) and leak towards the neutron band. In the traditional phase space this causes a distortion in both the y-axis, log$_{10}$(S2/S1), and x-axis, S1, such that:
 $$
[\mathrm{log_{10}(S2/S1) ,~ S1}]  \rightarrow [\mathrm{log_{10}(S2/(S1+S1')) ,~ (S1+S1')}]  
$$
In the new space only the y-axis is affected:
$$
[\mathrm{log_{10}(E1/E2) ,~E2}]  \rightarrow [\mathrm{log_{10}((E1+E1')/E2) ,~E2}]  
$$
The result is that the MSSI events produce a flat distribution in the E2 phase space that is more readily characterised.  

The consequences of the new approach on the electron recoil rejection is being explored by developing a similar model of (S1, C2) observable signals from $\gamma$-rays. However it is clear that such an approach (translating to different phase spaces) cannot introduce any additional background events to a signal box - every event in any of the four phase spaces can be mapped to any other phase space.  The number of background events in one phase space remains exactly the same in any other. 

Finally, with regards any future positive signal, since the proposed method dramatically improves both the acceptance and the energy determination whilst maintaining discrimination, the WIMP mass may be determined more reliably than as presented in previous studies for the future large-scale detectors~\cite{Arisaka:2009,Arisaka:2011}, given appropriate statistics. A thorough exploration with this revised analysis will be presented in a separate article. 

\subsection{Conclusions}

Traditionally, nuclear recoil energy for WIMP searches by liquid Xe dual-phase TPCs has been estimated by means of the initial scintillation and not by the ionization charge. The secondary scintillation is exploited only as a discrimination parameter in log$_{10}$(S2/S1) vs. S1.  Alternative approaches that use only the S2 channel seriously damage the discrimination power and 3D event vertex information.  

This article has presented issues with the traditional approach, primarily due to the  fluctuations in S1 signal that do not allow reliable determination of the true recoil energy.  Additionally, the acceptance when defined in S1 does not translate to the assumed nuclear recoil energy, causing efficiency loss even towards the higher energy boundary of the signal box.

To address these problems, the new phase space of log$_{10}$(S1/C2) vs. C2 has been introduced using C2 as the superior energy estimator.  Crucially, the improved resolution can be exploited without loss of fiducialization (critical to liquid noble gas TPCs for self-shielding and event reconstruction) or discrimination (critical in rejecting single scatter electronic background).  

Finally, two energy estimators are defined (E1 from S1, and E2 from C2), and log$_{10}$(E1/E2) vs. E2 is introduced. Advantages of this approach may be summarized as follows:
\begin{enumerate}
\item The energy resolution from E2 is considerably enhanced relative to the E1, especially at low energies where the WIMP spectrum peaks.  This provides improvement in evaluating the WIMP energy spectrum against the background.
\item Lower energy thresholds and greater signal acceptance can be achieved without compromizing self-shielding or discrimination power.
\item Monte Carlo simulation models of neutron calibrations can be assessed in multiple phase spaces to address underlying assumptions and nuclear recoil efficiency can be uniquely extracted.  
\item For a given E2 bin, log$_{10}$(E1/E2) represents an electronic rejection parameter specific to that energy.
\item With multiple WIMP events observed the WIMP mass can be estimated with greater confidence.
\end{enumerate}

In conclusion, the proposed method of log$_{10}$(E1/E2) vs. E2 is more accurate and reliable than the traditional method of log$_{10}$(S2/S1) vs. S1.  E1 and E2 are presented as more accessible detector independent parameters that can be adopted within the dark matter search community and directly compared with other technologies.

\begin{acknowledgements}

We gratefully acknowledge support for this work in part by US DOE grant DE-FG02-91ER40662, and NSF grant PHY-0919363.

\end{acknowledgements}

\bibliography{MCPaper}

\end{document}